\documentclass[twocolumn,amssymb, nobibnotes, aps, prb]{revtex4-2}
\usepackage{amssymb}
\usepackage{appendix}
\usepackage{verbatim}
\usepackage{float}
\usepackage{color}
\usepackage{textcomp}
\usepackage{gensymb}
\usepackage{hyperref}
\usepackage{dsfont}
\hypersetup{
     colorlinks   = true,
     citecolor    = blue
}
\usepackage{siunitx}
\usepackage{graphicx}
\usepackage{dcolumn}
\usepackage{bm}
\usepackage{braket}
\usepackage{empheq}
\usepackage{nicefrac} 
\usepackage[dvipsnames]{xcolor}
\renewcommand{\vec}[1]{\bm{#1}}

\newcommand{\be}{\begin{equation}}
\newcommand{\ee}{\end{equation}}
\newcommand{\bea}{\begin{eqnarray}}
\newcommand{\eea}{\end{eqnarray}}

\def\nn{\nonumber}
\def\lb{\label}
\def\pref{\eqref}

\def\g{\gamma}

\def\tt{\hat \tau}

\begin{document}

\title{Band structure of twisted bilayer graphene on hexagonal boron nitride} 

\author{Tommaso Cea$^1$}
\email[]{These authors contributed equally}
\author{Pierre A. Pantale\'on$^1$}
\email[]{These authors contributed equally}
\author{Francisco Guinea$^{1,2}$}
\affiliation{$^1$Imdea Nanoscience, Faraday 9, 28015 Madrid, Spain}
\affiliation{$^2$ Donostia International Physics Center, Paseo Manuel de Lardizábal 4, 20018 San Sebastián, Spain}

\date{\today}

\begin{abstract}
The effect of an hexagonal boron nitride (hBN) layer close aligned with twisted bilayer graphene (TBG) is studied. At sufficiently low angles between twisted bilayer graphene and hBN, $\theta_{hBN} \lesssim 2^\circ$, the graphene electronic structure is strongly disturbed. The width of the low energy peak in the density of states changes from $W \sim 5 - 10$ meV for a decoupled system to $\sim 20 - 30$ meV. Spikes in the density of states due to van Hove singularities are smoothed out. We find that for a realistic combination of the twist angle in the TBG and the twist angle between the hBN and the graphene layer the system can be described using a single moir\'e unit cell.
\end{abstract}

\maketitle

{\it Introduction.}
The discovery of insulating behavior at integer filling and superconductivity in TBG~\cite{Cao2018,Cao2018_bis,Kim2017,Huang2018a,Yankowitz2019}
motivated a recent effort in the study of a wide class of van der Waals heterostructures,
displaying moir\'e patterns on length scales much larger than the lattice constant of their constituent layers. The periodicity induced by the moir\'e can affect sensitively the electronic structure of the material, giving rise to narrow, almost dispersionless, flat bands. The kinetic quenching may favor the interactions between the electrons,
paving the way for the appearance of strongly correlated phases.

So far, transport measurements on TBG have been performed with the sample
either encapsulated between two hBN clapping layers, see e.g.~\cite{Cao2018,Cao2018_bis,Xiaobo_nat19,Tomarken2019,Setal20,stepanov_cm19,Zondiner2019}, or suspended on a substrate of hBN~\cite{Xie2019,Wong2019}.
Recently, it has been observed~\cite{Singh_cm2020} that the presence of an additional layer of WS$_2$ between hBN and TBG stabilizes the SC phase in a range of angles wider than reported previously.
The presence of hBN breaks the in-plane two-fold rotational symmetry ($\mathcal{C}_2$), gapping out the Dirac crossing of monolayer graphene~\cite{Hunt2013,SL13,Amet2013,Getal14,Cetal14,Yankowitz2014,Wetal15,Jung2015,Lee_science2016,Wetal16,Yankowitz_nat2018,Zibrov_natphys18,Kim2018}. Placing TBG on top of hBN accounts for two coexisting moir\'e patterns: that induced by the mismatch between the lattice constants of hBN and graphene~\cite{Xue2011,Yankowitz2012b,Woods2014,Moon2014,San-Jose2014}, and that induced by the relative orientation between the two graphene layers of the TBG. If the hBN is aligned with the adjacent graphene layer and the relative twist between the two layers of graphene is close to $1^\circ$, then the two moires have very similar periods, $\sim 13$nm, even if they are not commensurate. This sensitively affects the band structure of TBG close to charge neutrality (CN). To a first approximation, it is expected that the breaking of inversion symmetry induced by the hBN layer gives rise to a gap at the Dirac point of the TBG, separating two flat conduction and valence bands, carrying opposite Chern numbers, ${\cal C}=\pm 1$~\cite{Zhang2019b,bultinck_cm2019}. This analysis may explain the observed anomalous Hall effect at the integer band filling, $\nu=3$~\cite{Setal20,Serlin2019,Sawinska2010b}. The existence of flat bands can lead to Chern insulators, with features similar to those found in the Quantum Hall Effect \citep{BKLCVZ19,KCBZV20}.


In the following, we assume that the twist angle, $\theta_{TBG}$, in the TBG is fixed to a value near a magic angle, and study the effect of a hBN layer as function of the angle between this layer and the neighboring graphene layer, $\theta_{hBN}$. The periodicities of the two moir\'e patterns described above are $L_{TBG} \approx d_G / \theta_{TBG}$ and $L_{G,hBN} \approx d_G / \sqrt{\theta_{hBN}^2 + ( d_{hBN} / d_G )^2}$, where $d_G$ and $d_{hBN}$ are the lattice constants of graphene and hBN. 
The overall structure resembles the arrangement in a twisted graphene trilayer, or in twisted graphite~\cite{Amorim2018,MRB19,Cea2019a}. The two moir\'e lattices define a generically incommensurate superstructure. 

Interestingly, a realistic choice of parameters allows us to define a single moir\'e unit cell for the whole system. This possibility permits an accurate study of the electronic properties. Interaction effects are included using the unrestricted Hartree Fock approximation~\cite{CG20}. We analyze the similarities and differences with the electronic structure of TBG decoupled from the substrate, and also with other graphitic systems which show narrow bands~\cite{PCBWG20}.

\begin{figure*}
\centering
\includegraphics[scale=0.25]{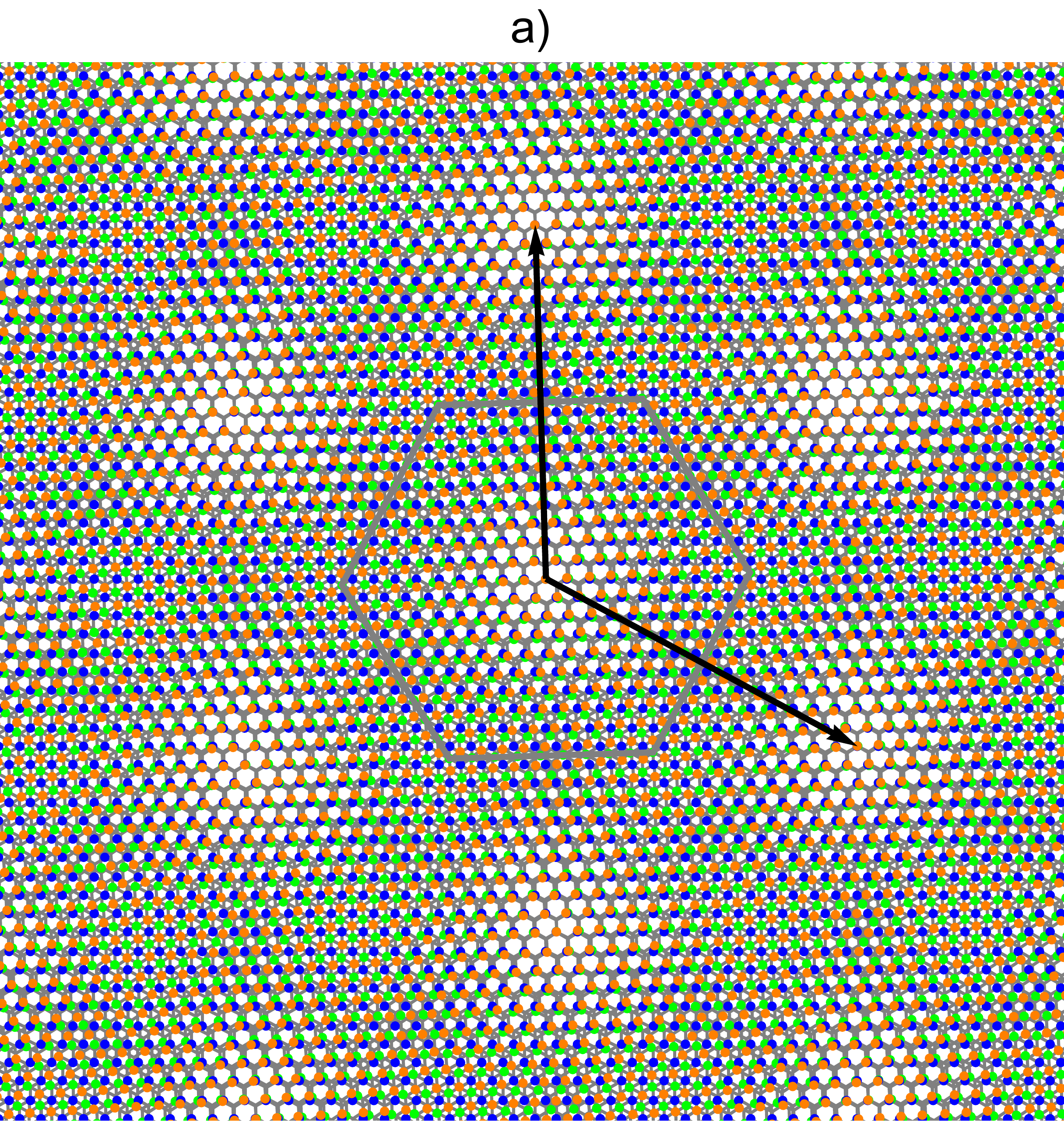}
\includegraphics[scale=0.25]{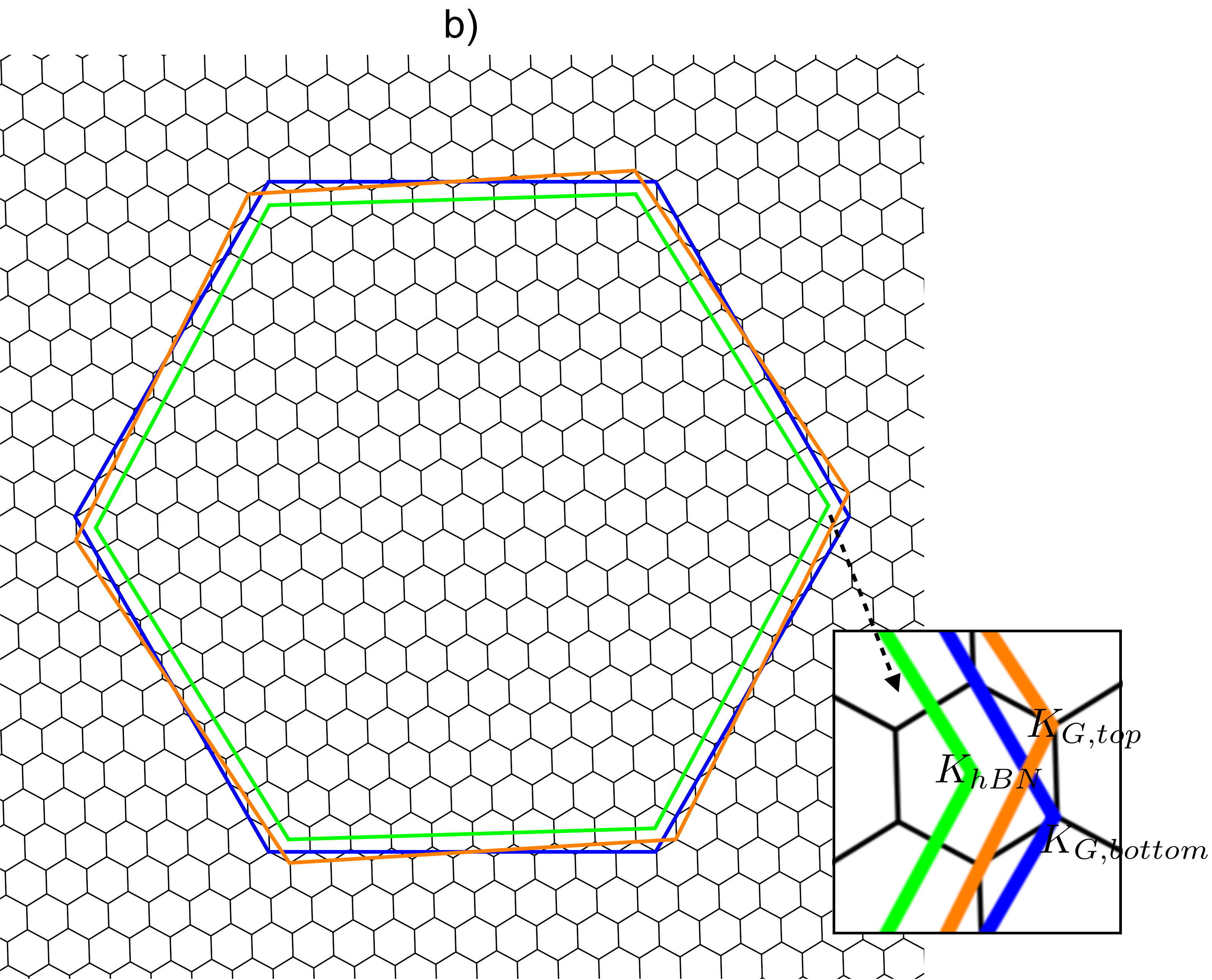}
\caption{a) Sketch of the moir\'e superlattice.
The blue and orange points represent the carbon atoms, while the green points refer to the substrate.
b) The large hexagons represent the BZs of the constituting layers. Their folding gives rise to the mini-BZs represented by the small black hexagons. In the inset: one side of the mini-BZ connects the corners of the BZs of each pair of layers.}
\label{sketch}
\end{figure*}
{\it The model.}
A sketch of the atomic arrangement to be considered, and of its Brillouin zone is shown in Fig.~[\ref{sketch}]. We assume, as in the continuum model for TBG~\cite{LopesDosSantos2007,Bistritzer2011} that one side of the hexagonal Brillouin zone of the superlattice connects the corners of the Brillouin zones of each pair of layers, see Fig. [\ref{sketch}](b). The positions of the corners of the three Brillouin zones are
\begin{align}
    \vec{K}_{G, top} &\approx \frac{4 \pi}{3 d_G} \left( \vec{n}_x + \theta_{TBG} \vec{n}_y \right) \nonumber \\
    \vec{K}_{G, bottom} &= \frac{4 \pi}{3 d_G} \vec{n}_x \nonumber \\
    \vec{K}_{hBN} &\approx \frac{4 \pi}{3 d_{hBN}} \left( \vec{n}_x - \theta_{hBN} \vec{n}_y \right) \approx \nonumber \\
    &\approx \frac{4 \pi}{3 d_{G}}  \left( 1 - \frac{d_{hBN}}{d_G} \vec{n}_x - \theta_{hBN} \vec{n}_y \right)
\label{bz}
\end{align}
 where $\vec{n}_x$ and $\vec{n}_y$ are unit vectors along the $x$ and $y$ axes, and we have expanded the exact expressions to lowest order, $\theta_{TBG} , \theta_{hBN} , d_{hBN} / d_G - 1 \ll 1$.
 In order for the two moires to have the same unit cell, we need the vectors $\vec{K}_{G, top} - \vec{K}_{G, bottom}$ and $\vec{K}_{G, top} - \vec{K}_{hBN}$ to have the same length, and to make an angle equal to $(2 \pi )/3$. These two conditions imply that:
 \begin{align}
     \theta_{TBG} &\approx \sqrt{\theta_{hBN}^2 + \left( \frac{d_{hBN}}{d_G} - 1 \right)^2 }, \nonumber \\
     \frac{\theta_{hBN}}{\theta_{TBG}} &= \frac{1}{2}.
     \label{comp}
 \end{align}
For a fixed value of $d_{hBN} / d_G$ these equations are satisfied when
\begin{align}
    \theta_{TBG} \approx 2 \theta_{hBN} \approx \frac{2 }{\sqrt{3}} \left( \frac{d_{hBN}}{d_G} - 1 \right).
\end{align}
For $d_G = 2.46$\AA, $d_{hBN} = 2.50$\AA, and $d_{hBN} / d_G - 1 \approx 0.017$ we obtain $\theta_{TBG} \simeq 1.05^\circ$. This number is reasonably close to the twist angles where TBG shows a non trivial phase  diagram. The twist of hBN, $\theta_{hBN}\approx 0.52^\circ$ is close to perfect alignment. The presence of a unique moir\'e pattern in hBN/TBG heterostructures is consistent with recent scanning-tunneling-microscopy (STM) maps~\cite{Wong2019}, where, in some samples,  only the moir\'e pattern identified by the TBG appears. 


 {\it Results. }
We model the Hamiltonian of the TBG within the low energy continuum model, see~\cite{LopesDosSantos2007,Bistritzer2011,LopesDosSantos2012,Koshino2018a}. The effect of the hBN is included by means of an effective periodic potential acting on the nearest graphene layer~\cite{Wallbank2013,San-Jose2014}.
In what follows we refer to the parametrization of such potential as given in the Ref.~\cite{Jung2017}.
A detailed description of the model is given in~\cite{si}. 

We first study the arrangement described by the angles $\theta_{TBG} = 1.05^\circ$ and $\theta_{hBN} = 0.52^\circ$, where a single moir\'e unit cell describes the system, as shown in Fig.~[\ref{sketch}]. Note that three different stacking configurations of the twisted system can be identified, see~\cite{si}, Figs.~[\ref{moire_cells}] and [\ref{stackings}]. The three stackings differ in the relative arrangement of layers which are second nearest neighbors. As shown below, the three geometries lead to different electronic structures. 

\begin{figure*}
\centering
\includegraphics[scale=0.45]{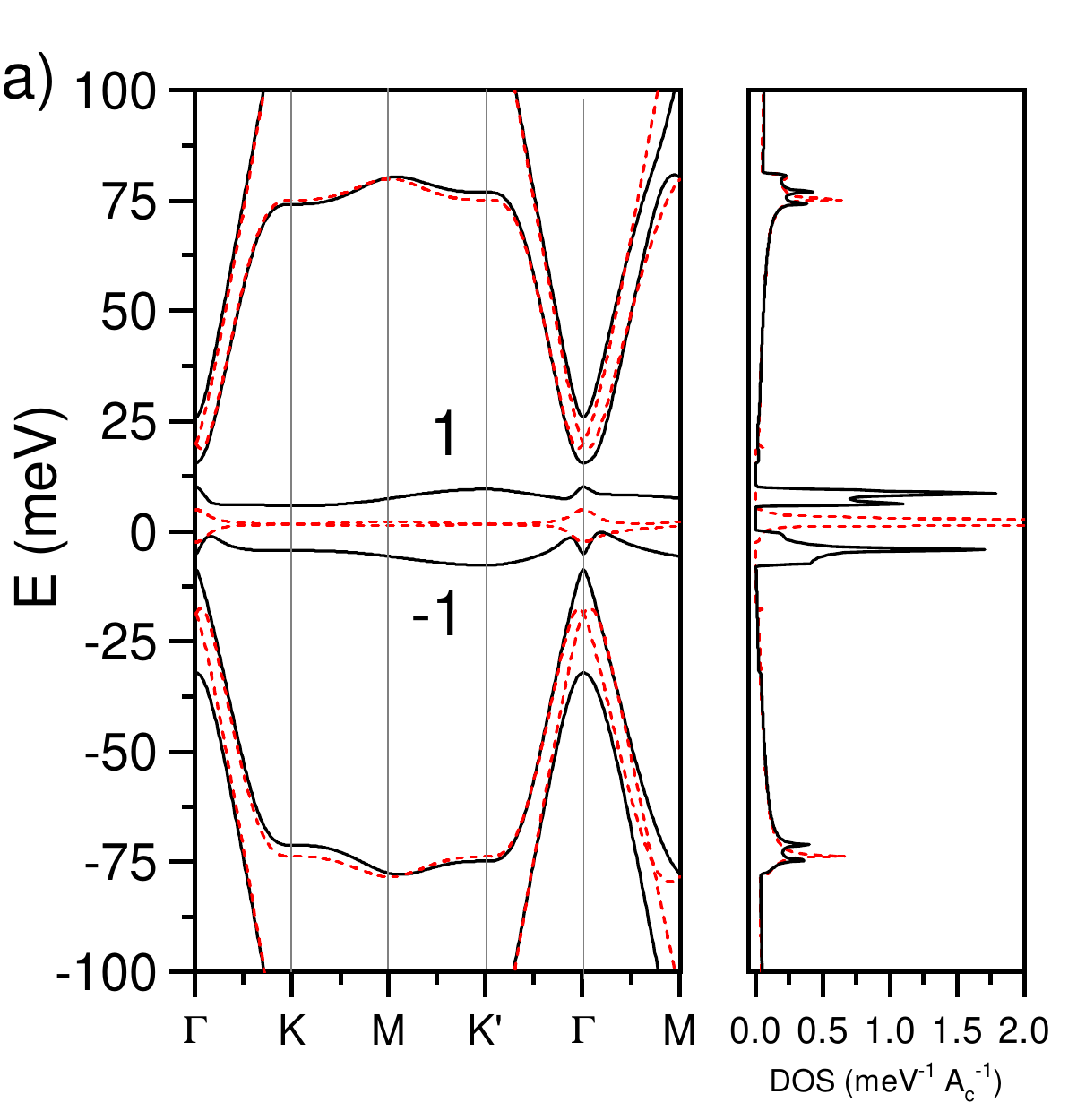}
\includegraphics[scale=0.45]{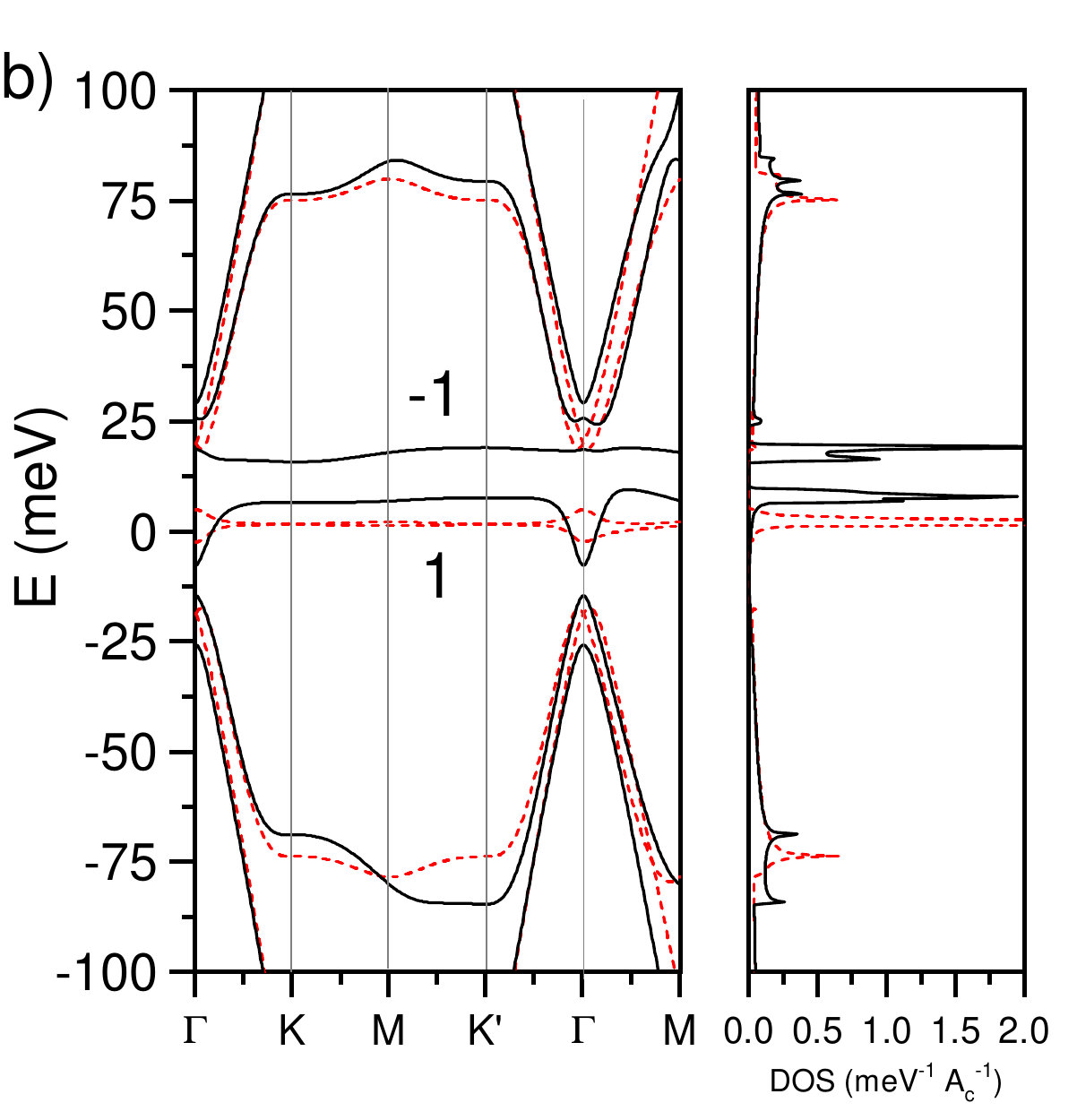}
\includegraphics[scale=0.45]{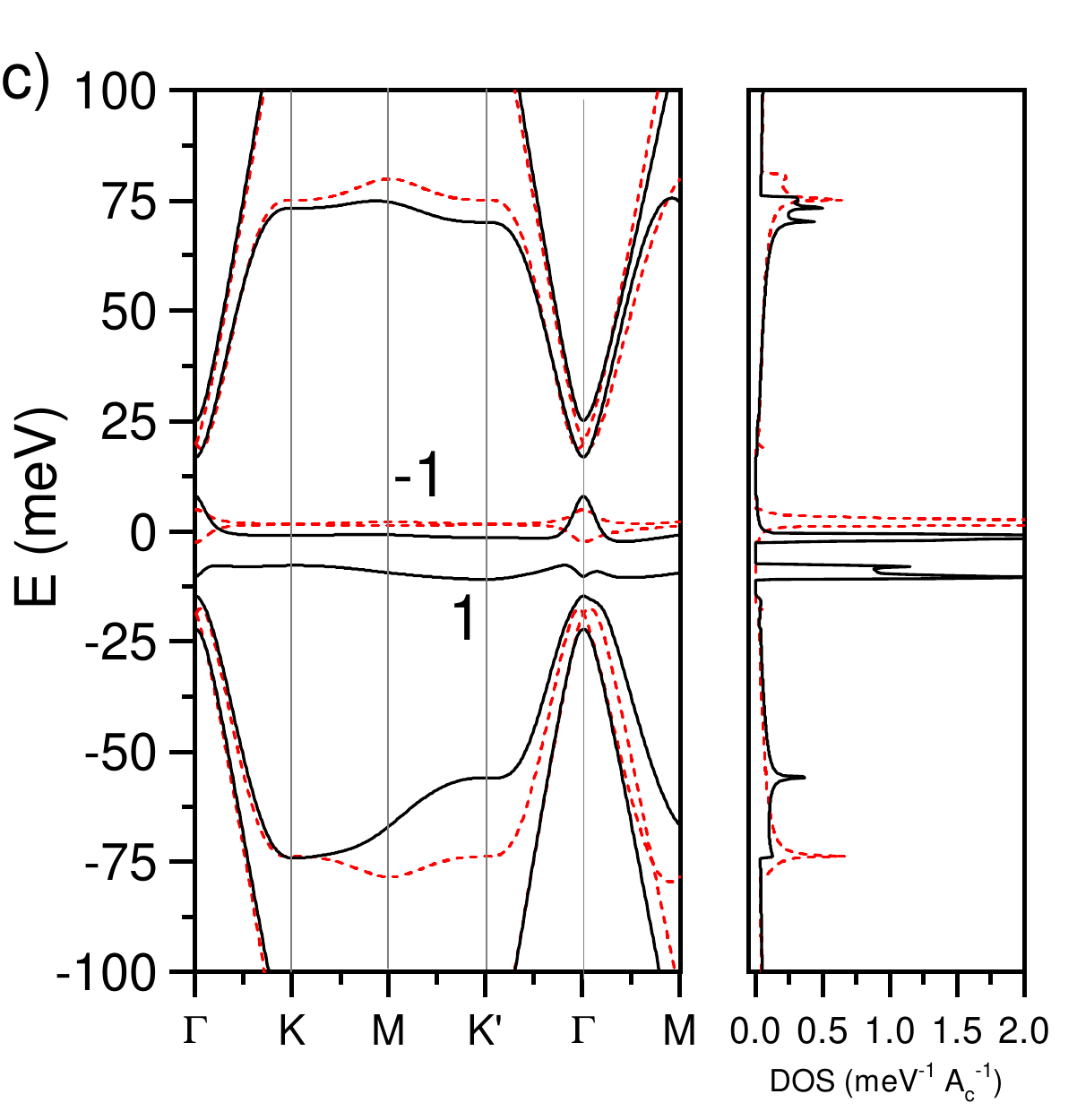}
\caption{Band Structure of TBG on top of hBN in three different configurations, a) AAA, b) CAA and c) BAA stacking, respectively. The relative twist between the two graphene layers is $\theta_{TBG}=1.05^\circ$. The red dashed lines indicate energy bands and DOS, respectively, of the TBG decoupled from the substrate. The bands are computed in the $K$ valley. The valley Chern numbers for the corresponding bands are indicated.    
}
\label{Fig: BandAA}
\end{figure*}

\begin{figure}
\centering
\includegraphics[scale=0.55]{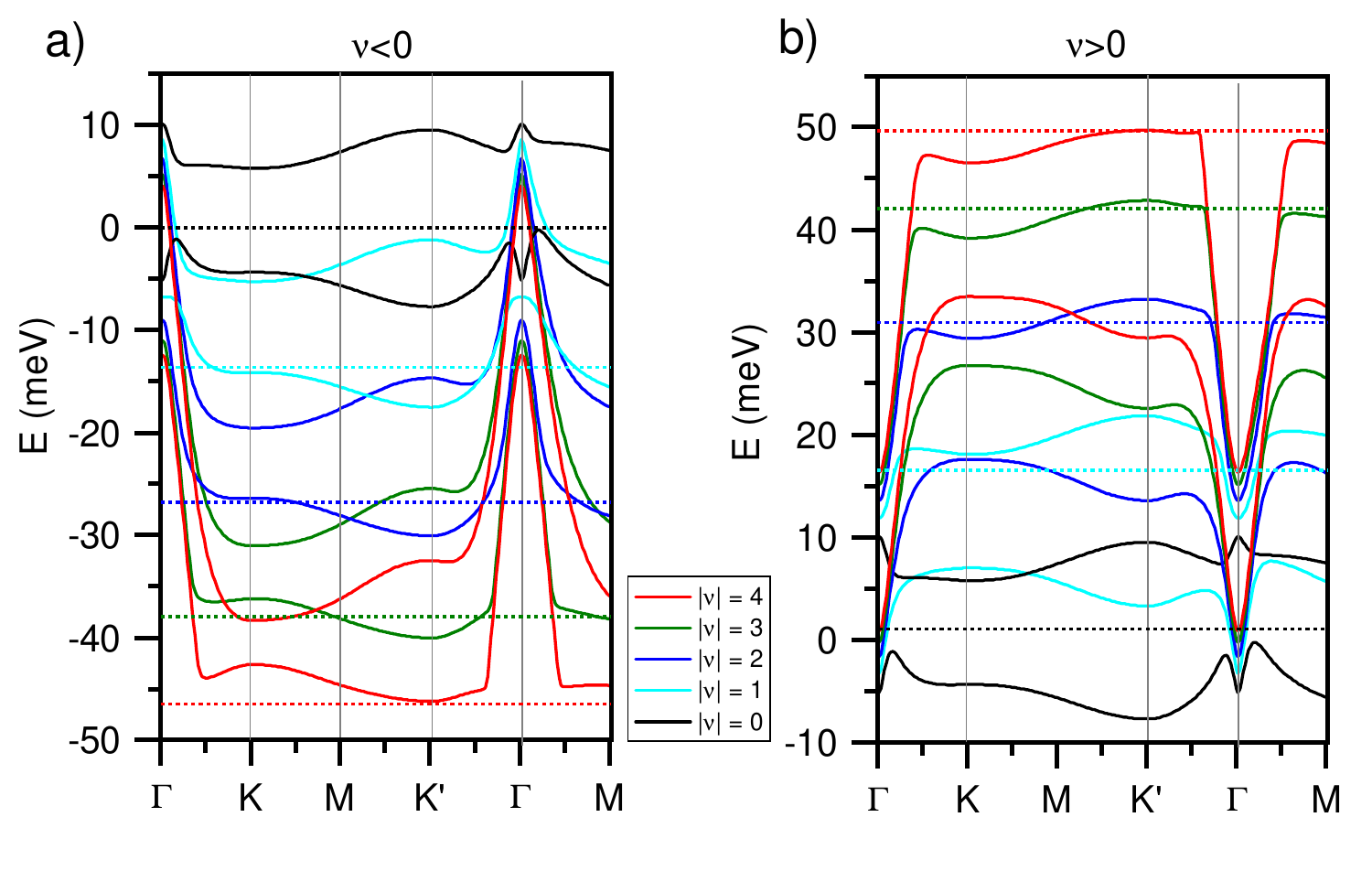}
\caption{Self-consistent bands of AAA stacked hBN/TBG obtained for a twist angle of $\theta_{TBG}=1.05^\circ$ with a) negative and b) positive filling fractions. The horizontal dashed lines represent the Fermi energies.}
\label{Fig: BandHartree}
\end{figure}

\begin{figure}
\centering
\includegraphics[scale=0.65]{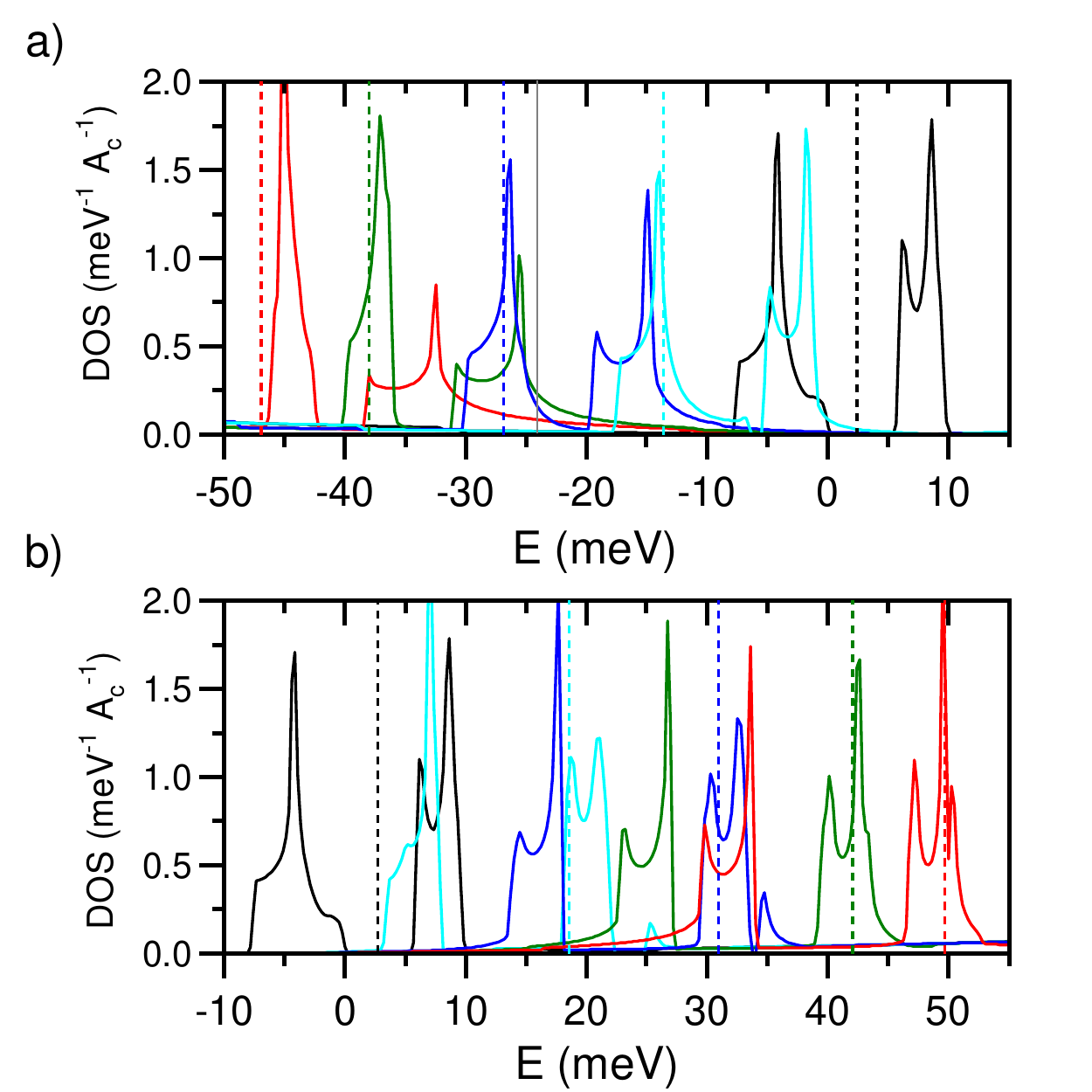}

\caption{DOS for each of the results shown in Fig.~\ref{Fig: BandHartree} for a filling a) $\nu<0$ and b) $\nu>0$, color coded as in those figures. The vertical dashed lines represent the Fermi energy for each case. }
\label{Fig: BandDOS}
\end{figure}

The band structure and density of the states (DOS) per unit cell of the hBN/TBG are shown in the Fig. [\ref{Fig: BandAA}] (black lines), and compared to that of the TBG (red dashed lines). 
$A_{c}=\sqrt{3}L_{TBG}^2/2$ is the area of the moir\'e unit cell. The presence of the substrate strongly affects the spectrum of the TBG. The flat bands at CN of the TBG become  dispersive in the hBN/TBG stack, acquiring a finite bandwidth of $\sim 6-8$ meV, which is almost twice that of the TBG. As a consequence, the peak in the DOS of the TBG at CN is strongly smoothed and also split in the hBN/TBG, giving rise to an insulating structure with a small band gap, which is due to the breaking of $\mathcal{C}_2$ induced by the hBN layer. The effect of a self consistent Hartree potential away from CN is shown in Fig. [\ref{Fig: BandHartree}] and [\ref{Fig: BandDOS}]. The Chern numbers of these bands are shown in~\cite{si}. It is worth mentioning that Chern numbers of up to ${\cal C} = 3$ are obtained. The effect of the exchange term at the neutrality point is analysed in~\cite{si}. As in the absence of a substrate~\cite{Guinea2018a,Cea2019}, the bands are significantly distorted by the Hartree potential.

 Generic values of the twist angle between hBN and TBG cannot be described by a simple moir\'e unit cell. This is, for example, the case for perfect alignment: $\theta_{hBN}=0^\circ$.  Because of the lack of commensuration, we cannot define a crystal momentum. In order to study the energy spectrum, we project the perturbation induced by the hBN on the low energy states of the TBG and solve a dual lattice in the reciprocal space of the TBG, as detailed in~\cite{si}. The scheme follows closely continuum models for a graphene monolayer on hBN, where the perturbation due to the hBN layer is projected onto the graphene Dirac cone. An infinite number of minibands emerge, induced by the periodicity of the potential due to the hBN layer. In a similar manner, the TBG bands are replicated and coupled in our calculation.
A similar method has been recently used in the Ref.~\cite{MKS19} for studying the quasi-crystalline electronic spectrum of the non-commensurate 30$^\circ$-TBG.

\begin{figure*}
\centering
\includegraphics[scale=0.3]{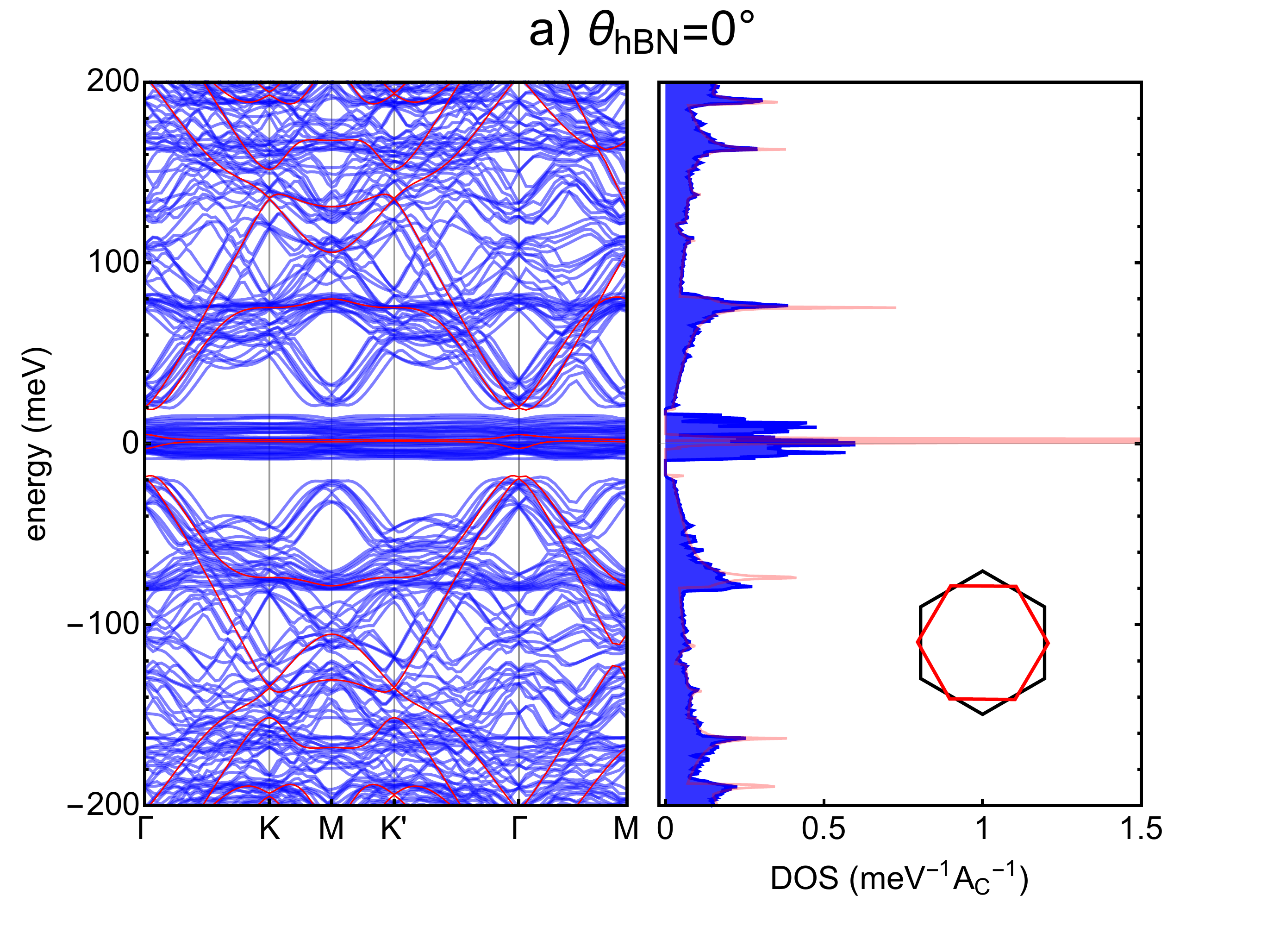}
\includegraphics[scale=0.3]{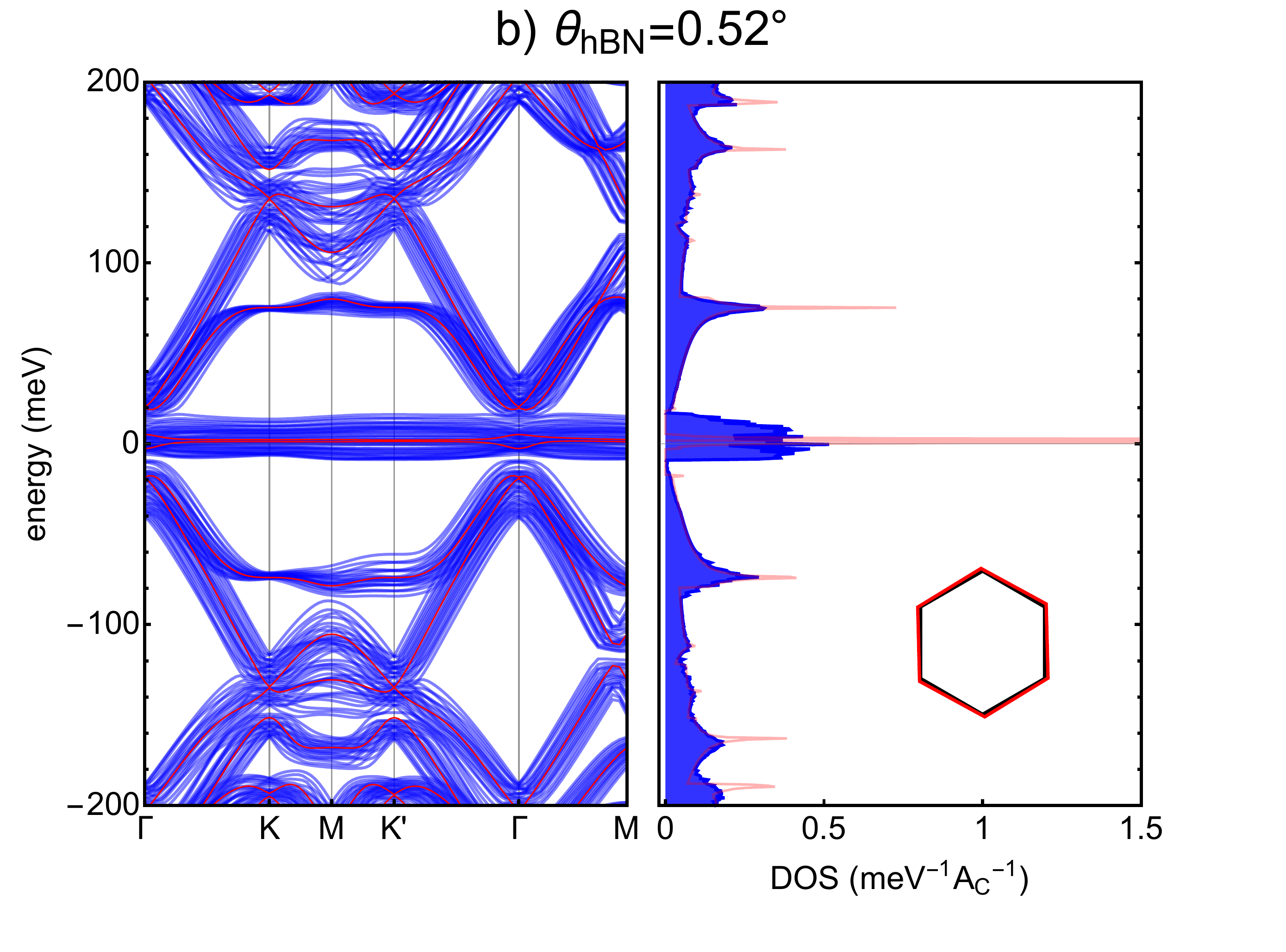}\\
\includegraphics[scale=0.3]{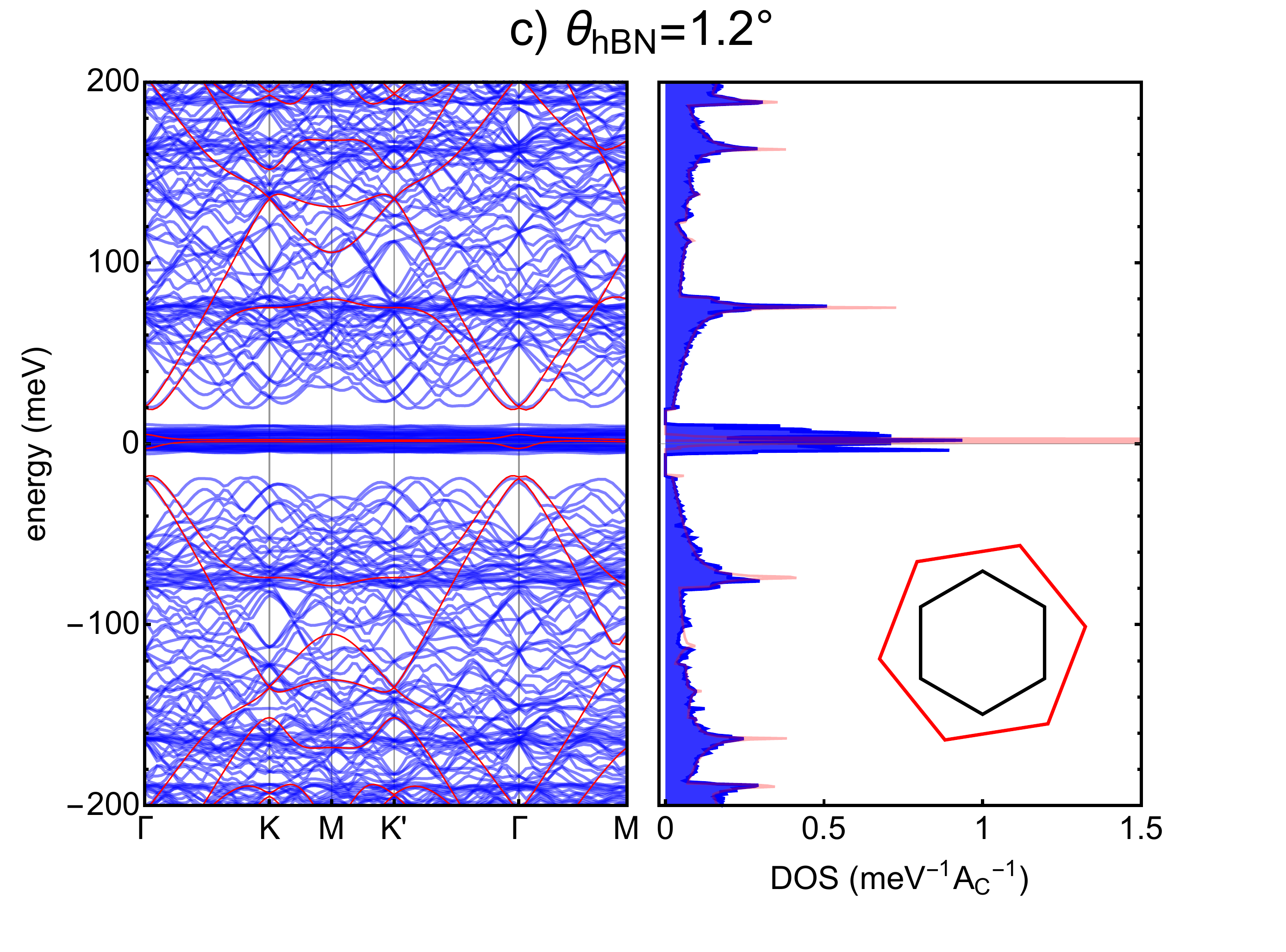}
\includegraphics[scale=0.3]{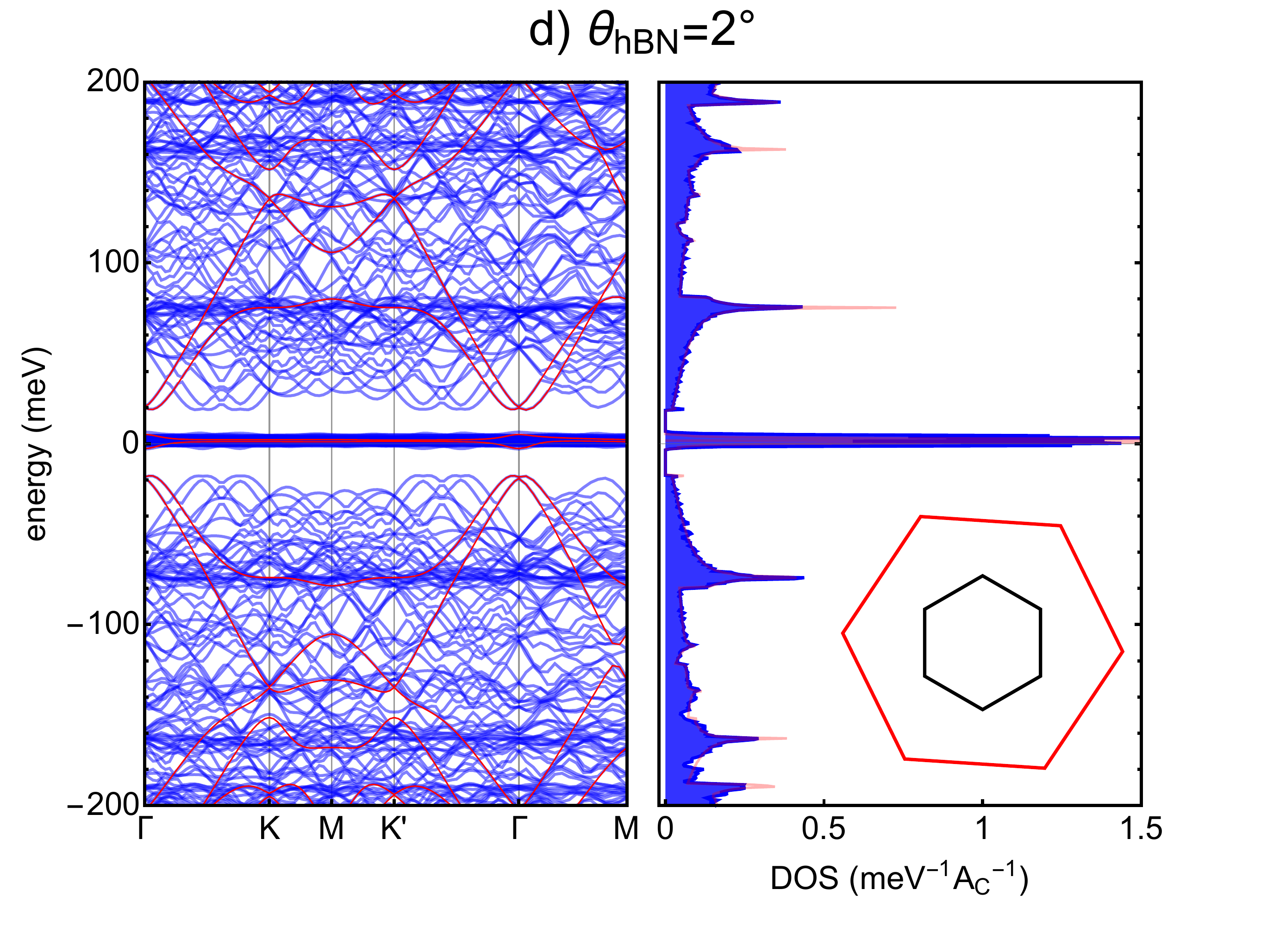}
\caption{Quasi-band structure and DOS of the non-commensurate hBN/TBG.
The black and red hexagons show the two different BZs of the TBG and of the hBN/G, respectively.
The red lines refer to the band structure and the DOS of the unperturbed TBG.}
\label{non_commensurate_figs}
\end{figure*}

The  quasi-band structures, obtained by varying the momentum $\vec{k}$ in the BZ of the TBG, and the DOS are shown in the Fig. [\ref{non_commensurate_figs}] for different orientations between the hBN and graphene, and $\theta_{TBG}=1.05^\circ$.
The black and red hexagons show the two different BZs of the TBG and of the hBN/G, respectively.
The red lines refer to the band structure and the DOS of the unperturbed TBG, that also includes the long-wavelength staggered potential induced by the hBN, weakly breaking the $\mathcal{C}_2$ symmetry of graphene (see \cite{si}).
As is evident, the hBN strongly affects the spectrum close to CN at small angles, $\theta_{hBN}<1^\circ$, where the moir\'e identified by the TBG and that identified by the substrate of hBN have similar periods.
In contrast to the narrow bands of the TBG, the hBN/TBG exhibits a broad structure, with a bandwidth of approximatively 30meV, which can even overlap with the higher energy bands.
This is for example the case of $\theta_{hBN}=0.52^\circ$, Fig. [\ref{non_commensurate_figs}](b), which is close to commensuration. 
 In addition, the band broadening at CN lowers the DOS of the hBN/TBG as compared to the sharp van Hove singularity of the TBG, that has been cut out of the energy scale in the central panels of Fig. [\ref{non_commensurate_figs}].
 In general, the high energy spectrum is barely affected by the hBN.
 Upon increasing $\theta_{hBN}$, the bandwidth at CN gradually shrinks, while the DOS gains intensity and the central bands further separate from the rest of the spectrum. At $\theta_{hBN}=2^\circ$ the effect of the hBN is almost completely negligible and we recover the narrow band feature of the TBG, except for a constant gap.
 Even though we are using a small value of the staggered potential, $\Delta=3.62$meV, as given by the Ref.~\cite{Jung2017}, we checked that our results remain valid in a wide range of values of $\Delta$. The details are shown in~\cite{si}, where we report the case of $\Delta=40$meV. 
 
 Finally, we show plots of the changes in the charge density distribution induced by the substrate in~\cite{si}. The sixfold symmetry of the TBG is reduced to threefold, in the case where the two moirés coincide. We expect a lower symmetry in the general case. These results are in agreement with the reduced symmetry observed in STM experiments~\cite{Jetal19}.

{\it Conclusions.}
We have analyzed the effect of a nearly aligned hBN substrate on the low energy bands of a twisted graphene bilayer. We find that at large enough angles between the TBG and the next hBN layer, $\theta_{hBN} \gtrsim 2^\circ$, the TBG is effectively decoupled from the substrate. Only those effects which do not average to zero over the
hBN/G moir\'e unit cell, such as a finite gap due to the lack of inversion symmetry, survive. 

On the other hand, perturbations of zero average over the unit cell change significantly the electronic structure. The narrow peak in the DOS associated to the narrow bands of the TBG broadens appreciably. The van Hove singularities are smoothed out, and the width of the peak near the Dirac energy increases from $ W \sim 5- 10$ meV for a decoupled TBG to $W \gtrsim 20 -30$ meV for a well aligned hBN/TBG stack. This peak overlaps with higher energy bands. In addition, the gap expected from the lack of inversion symmetry becomes filled with states which arise from the non uniform part of the perturbation due to the substrate.

{\it Acknowledgements.}
This work was supported by funding from the European Commision, under the Graphene Flagship, Core 3, grant number 881603, and by the grants NMAT2D (Comunidad de Madrid, Spain),  SprQuMat and SEV-2016-0686, (Ministerio de Ciencia e Innovación, Spain).

This paper, main ideas, theory and figures have being developed during the COVID-19 lockdown.
\bibliography{Referencias}


\clearpage
\onecolumngrid

\setcounter{equation}{0}
\setcounter{figure}{0}
\setcounter{table}{0}
\setcounter{page}{1}
\makeatletter
\renewcommand{\theequation}{S\arabic{equation}}
\renewcommand{\thefigure}{S\arabic{figure}}

\begin{center}
\Large Supplementary information for \\ Band structure of twisted bilayer graphene on hexagonal boron nitride
\end{center}

\section{Geometry of the superlattice}
\begin{figure}
\centering
\includegraphics[width=3.5in]{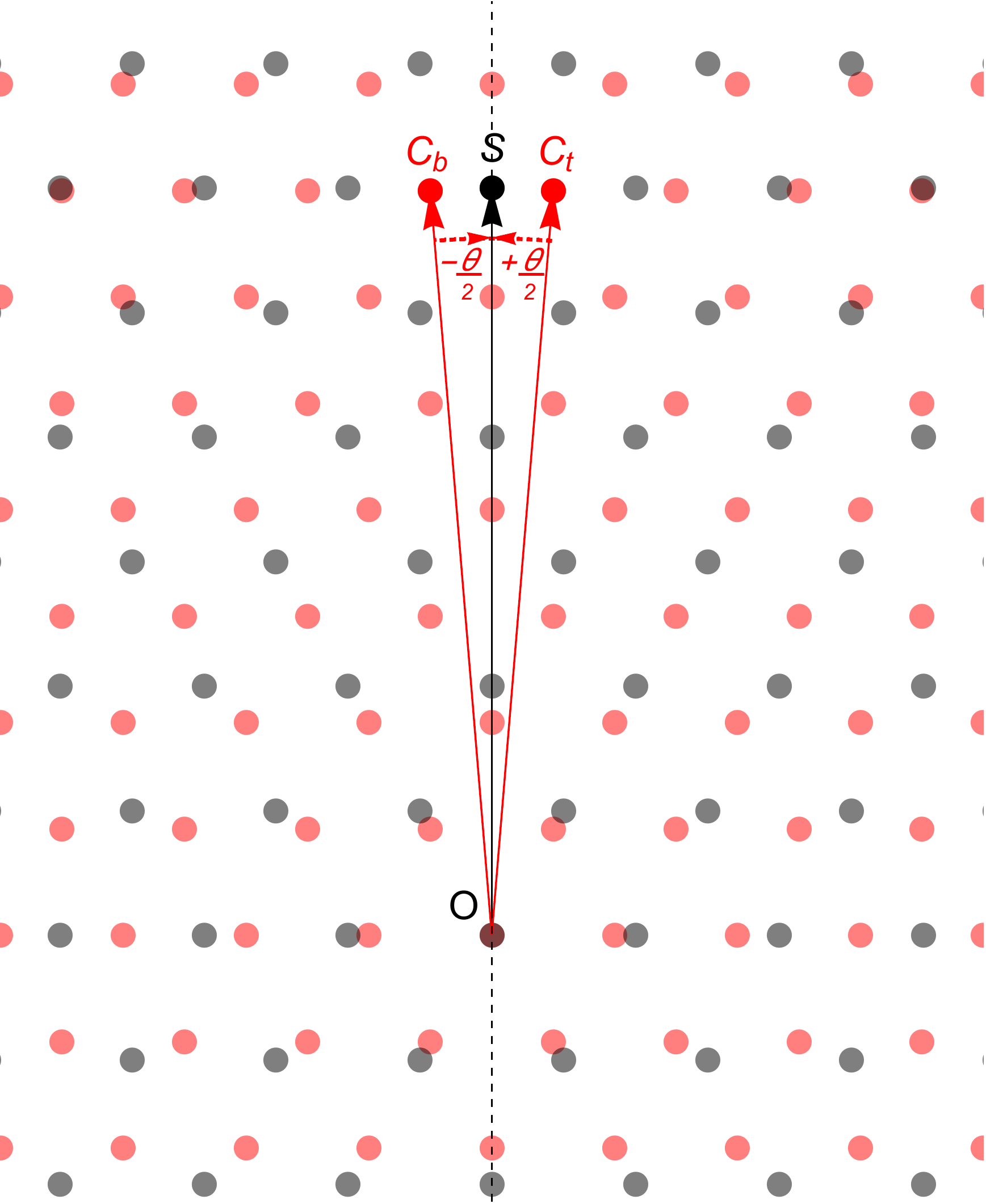}
\caption{
Sketch of the aligned Bravais lattices of graphene (red points) and the substrate (black points). $\theta$ is the angle identifying the TBG.
Keeping the substrate fixed with respect to the aligned configuration and
rotating the bottom (top) graphene layer by $-\theta/2$($+\theta/2$) around $O$,
moves the atoms $C_b$ and $C_t$ to the position occupied by $S$,
preserving the moir\'e superlattice of the TBG.}
\label{scheme_superlattice}
\end{figure}

We consider the heterostructure obtained by placing the TBG on top of a substrate with an hexagonal atomic structure and lattice constant $d_s$.
Let $\vec{d}_{G,1}=d_G\left(1,0\right)$, $\vec{d}_{G,2}=d_G\left(1/2,\sqrt{3}/2\right)$ be the primitive vectors of the Bravais lattice of the two unrotated layers of graphene, with $d_G=2.46${\AA}, and
$\vec{d}_{s,1}=d_s\left(1,0\right)$, $\vec{d}_{s,2}=d_s\left(1/2,\sqrt{3}/2\right)$ those of the substrate.
We assume that at the origin, $O=(0,0)$, the atomic positions of the three Bravais lattices coincide. Then, given $i,l$ integer numbers, we consider the atomic positions:
\begin{subequations}
\bea
C_b=(-i-1)\vec{d}_{G,1}+(2i+1)\vec{d}_{G,2}=
d_G\left(-1/2,(2i+1)\sqrt{3}/2\right),
\eea
of the bottom graphene layer,
\bea
C_t&=&-i\vec{d}_{G,1}+(2i+1)\vec{d}_{G,2}=
d_G\left(1/2,(2i+1)\sqrt{3}/2\right),
\eea
of the top graphene layer, and
\bea
S&=&-l\vec{d}_{s,1}+2l\vec{d}_{s,2}=d_s\left(0,l\sqrt{3}\right),
\eea
of the substrate.
\end{subequations}
A sketch of the aligned Bravais lattices,
with the atomic positions $C_b$, $C_t$ and $S$, is shown in Fig. \ref{scheme_superlattice}, where we set: $i=l=3$, for the sake of simplicity.
The moir\'e superlattice of the TBG is obtained by rotating the bottom graphene layer by $-\theta/2$ around $O$, and the top graphene layer by $+\theta/2$, where
$\theta=\cos^{-1}\left(\frac{3i^2+3i+1/2}{3i^2+3i+1}\right)$.
This rotation moves the atoms $C_b$ and $C_t$ to the same position, lying on the vertical axis. The period of the moir\'e is: $L=d_G\sqrt{3i^2+3i+1}=\frac{d_G}{2\sin\left(\theta/2\right)}$,
where $L=\left| \overrightarrow{OC_b}\right|=\left| \overrightarrow{OC_t}\right|$.
Keeping the substrate fixed with respect to the aligned configuration,
the moir\'e superlattice of the TBG is preserved by the presence of the substrate if:
\bea\label{comm_cond}
\left | \overrightarrow{OS} \right|=L\quad\Leftrightarrow
d_s=d_G\frac{\sqrt{3i^2+3i+1}}{\sqrt{3}l}.
\eea
This condition indeed implies that the rotation of the two graphene layers moves the atoms $C_b$ and $C_t$ to the position occupied by $S$.

The rhs of the Eq. \pref{comm_cond} expresses $d_s$ as a function of two integer numbers, $i$ and $l$, setting a sufficient condition for the moir\'e of the TBG to persist in the presence of the substrate. The relative twist between the substrate and each of the two graphene layers is: $\theta_s=\theta/2$. In particular, for $i=l=31$, the Eq. \pref{comm_cond} gives: $d_s\simeq 2.50${\AA}, which is actually the value of the lattice constant of hBN. The corresponding angle is: $\theta\simeq1.05^\circ$, with $L\simeq13.4$nm.

\section{The continuum model of the TBG}
\begin{figure*}
 \centering
\includegraphics[width=3.5in]{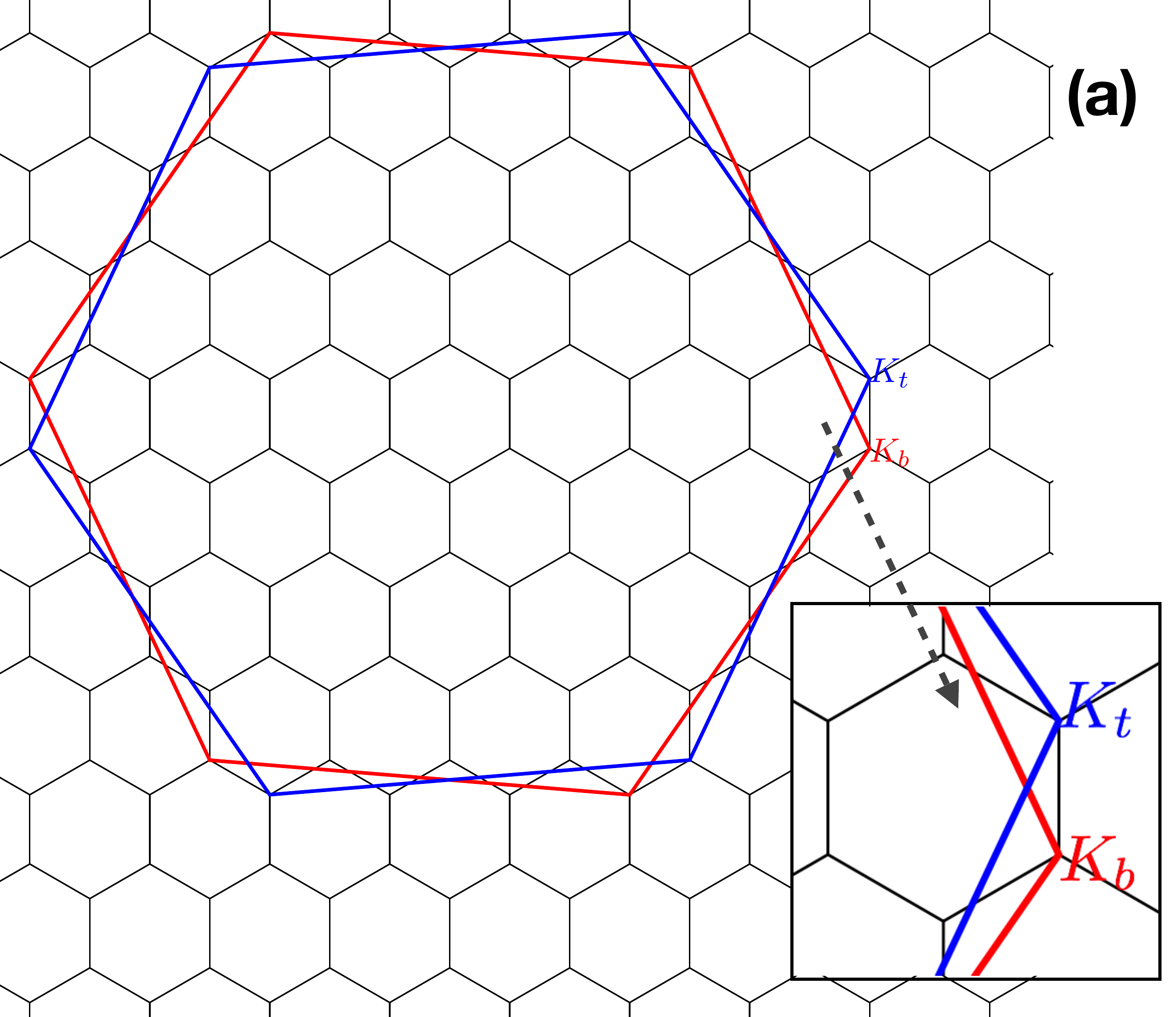}
\includegraphics[width=2.5in]{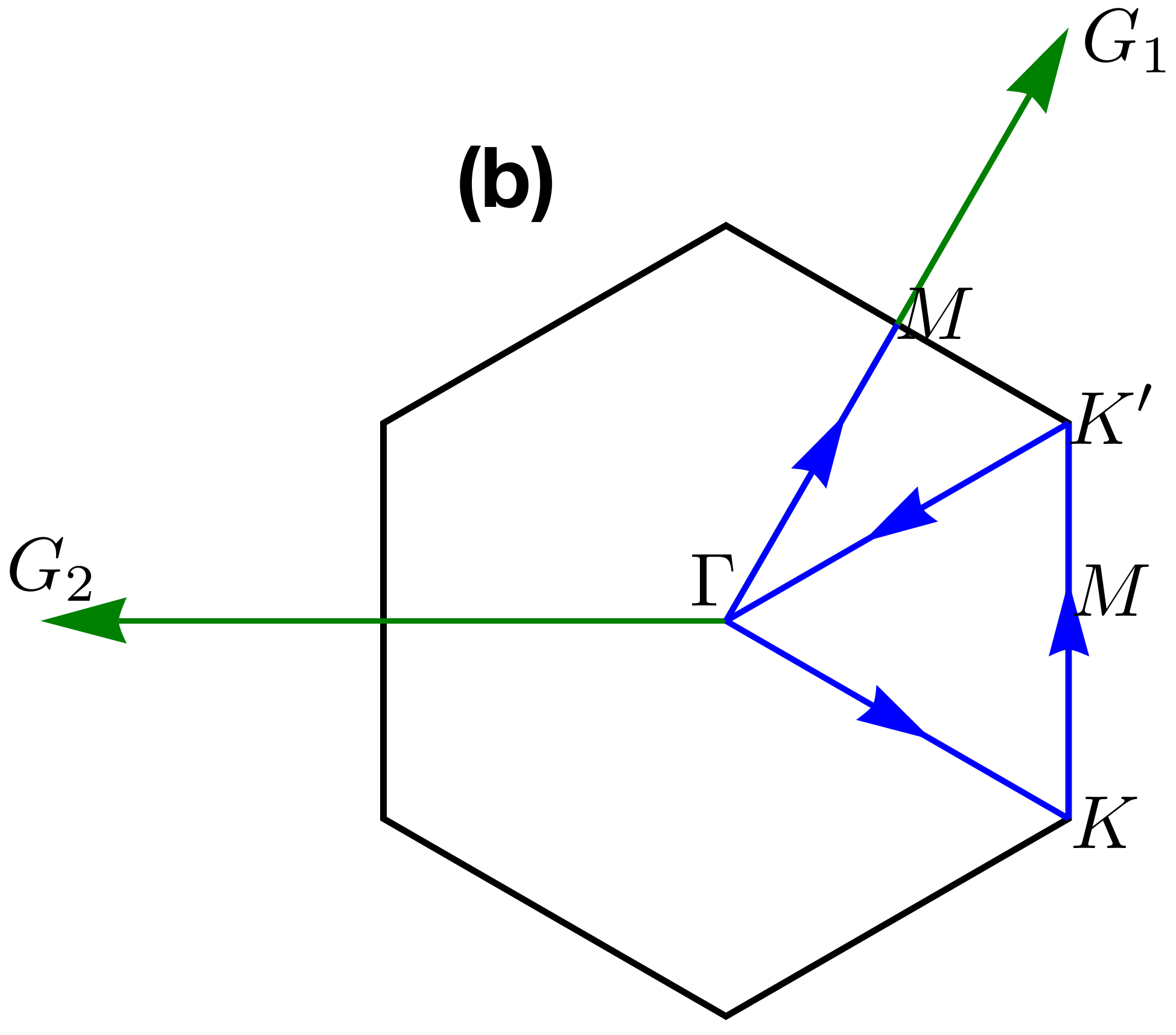}
\caption{
(a) Folding of the BZs of the twisted graphene monolayers.
The BZ of the bottom layer (red hexagon) is rotated by $-\theta/2$, while that of the top layer (blue hexagon) by $\theta/2$. The small black hexagons represent the mini-BZs forming the reciprocal moir\'e lattice. In the inset: $K_{b,t}$ are the Dirac points of the twisted monolayers, which identify the corners of the mini-BZ. (b) mini-BZ. $\vec{G}_1$ and $\vec{G}_2$ are the two basis vectors of the reciprocal lattice. The blue line shows the high symmetry path in the mini-BZ used to compute the bands shown in the main text.}
\label{BZ}
\end{figure*}

We describe the TBG within the low energy continuum model
considered in  Refs.\cite{LopesDosSantos2007,Bistritzer2011,LopesDosSantos2012,Koshino2018a},
which is meaningful for sufficiently small angles,
so that an approximatively commensurate structure can be defined for any twist.
The moir\'e mini-BZ, resulting from the folding of the two BZs of each monolayer
(see Fig.\ref{BZ}(a)),
is generated by the two reciprocal lattice vectors:
\begin{equation}
\vec{G}_1=2\pi(1/\sqrt{3},1)/L\text{ and } \vec{G}_2=4\pi(-1/\sqrt{3},0)/L,
\end{equation}
 shown in green in Fig.~\ref{BZ}(b).

For small angles of rotation the coupling between different valleys of the two monolayers
can be safely neglected, as the interlayer hopping has a long wavelength modulation. In what follows we describe the model for the behavior of the two $K$-valleys of the twisted monolayers, where $K=4\pi(1,0)/(3d_G)$ is the Dirac point of the unrotated monolayer graphene.
The case corresponding to the opposite valleys, at $K'=-K$, directly follows from time reversal symmetry, by inverting: $\vec{k}\rightarrow-\vec{k}$.

The Hamiltonian of the TBG is a $4\times4$ matrix, with entries: $\left(A_b,B_b,A_t,B_t\right)$,
where $A,B$ denote the sub-lattice and $b,t$ refer to the bottom and top layer, respectively.
Without loss of generality, we assume that in the aligned configuration, at $\theta=0$,
the two layers are $AA$-stacked.
In the continuum limit, the effective Hamiltonian of the TBG can be generally written as\cite{LopesDosSantos2007,Bistritzer2011,LopesDosSantos2012,Koshino2018a}:
\bea\lb{HTBG}
H_\text{TBG}=
\begin{pmatrix}
H_b&U(\vec{r})\\U^\dagger(\vec{r})&H_t
\end{pmatrix},
\eea
where:
\bea
H_l=\hbar v_F \left(-i\vec{\nabla}-K_l\right)\cdot
\vec{\tau}_{\theta_l}
\eea
is the Dirac Hamiltonian for the layer $l=b,t$, $v_F=\sqrt{3}td_G/(2\hbar)$ is the Fermi velocity, $t$ is the hopping amplitude between localized $p_z$ orbitals at nearest neighbors carbon atoms, $\theta_{b,t}=\mp\theta/2$, $\vec{\tau}_{\theta_l}=e^{i\tau_z\theta_l/2}\left(\tau_x,\tau_y\right)e^{-i\tau_z\theta_l/2}$,
$\tau_i$ are the Pauli matrices, and $K_l=4\pi\left(\cos\theta_l,\sin\theta_l\right)/(3d_G)$
are the Dirac points of the two twisted monolayers, which identify the corners of the mini-BZ shown in Fig.~\ref{BZ}. $U(\vec{r})$ is the interlayer potential, which is a periodic function in the moir\'e unit cell. In the limit of small angles, its leading harmonic expansion is determined by only three reciprocal lattice vectors\cite{LopesDosSantos2007}:
$U(\vec{r})=U(0)+U\left(-\vec{G}_1\right)e^{-i\vec{G}_1\cdot\vec{r}}+
U\left(-\vec{G}_1-\vec{G}_2\right)e^{-i\left(\vec{G}_1+\vec{G}_2\right)\cdot\vec{r}}$,
where the amplitudes $U\left(\vec{G}\right)$ are given by:
\bea
U(0)=\begin{pmatrix}g_1&g_2\\g_2&g_1\end{pmatrix}\quad,\quad
U\left(-\vec{G}_1\right)=\begin{pmatrix}g_1&g_2e^{-2i\pi/3}\\g_2e^{2i\pi/3}&g_1\end{pmatrix}
\quad,\quad
U\left(-\vec{G}_1-\vec{G}_2\right)&=&\begin{pmatrix}g_1&g_2e^{2i\pi/3}\\g_2e^{-2i\pi/3}&g_1\end{pmatrix}.
\eea
In the following we adopt the parametrization of the TBG given in the Ref.\cite{Koshino2018a}: $\hbar v_F/d_G=2.1354$eV, $g_1=0.0797$eV and $g_2=0.0975$eV. The difference between $g_1$ and $g_2$, as described in\cite{Koshino2018a}, accounts for the inhomogeneous interlayer distance, which is minimum in the $AB/BA$ regions and maximum in the $AA$ ones, or it can be seen as a model of a more complete treatment of lattice relaxation\cite{Guinea2019}. The Hamiltonian of Eq.~\pref{HTBG} then hybridizes states of the bottom layer with momentum $\vec{k}$ close to the Dirac point with the states of the top layer with momenta: $\vec{k},\vec{k}+\vec{G}_1,\vec{k}+\vec{G}_1+\vec{G}_2$.

The Hamiltonian of the Eq. \pref{HTBG} is diagonalized by Bloch eigenfunctions satisfying periodic boundary conditions in a region of space, $\Omega$, containing a large number of moir\'e unit cells:
\bea\label{Bloch_waves}
\ket{m,\vec{k}}=\frac{1}{\sqrt{\Omega}}\int_\Omega\,d^2\vec{r}\sum_{\vec{G}a}\phi_{m,\vec{k},a}\left(\vec{G}\right)e^{i\left(\vec{k}+\vec{G}\right)\cdot\vec{r}}\ket{\vec{r},a},
\eea
where $m$ is the band index, $\vec{k}$ is the momentum in the mini-BZ,
$\vec{G}=n_1\vec{G}_1+n_2\vec{G}_2$ are reciprocal lattice vectors, with $n_1,n_2$ integers,
$a$ is the sub-lattice/layer index
and $\phi_{m,\vec{k},a}\left(\vec{G}\right)$ are numerical eigenvectors.
Upon varying $\vec{k}$, the eigenvalue $E_m(\vec{k})$ corresponding to $\ket{m,\vec{k}}$ defines the $m$-th Bloch band.

Note that the $\phi$'s can be chosen within a gauge degree of freedom.
Given the gauge $\mathcal{U}$, the mapping between equivalent points of the reciprocal space is defined according to:
\begin{subequations}\label{gauge_eqs}
\bea\label{gauge_phi}
\phi_{m,\vec{k}+\vec{G}_0,a}\left(\vec{G}\right)=
\sum_n \mathcal{U}_{mn,\vec{k}}\left(\vec{G}_0\right)
\phi_{n,\vec{k},a}\left(\vec{G}+\vec{G}_0\right),
\eea
where $ \mathcal{U}_{\vec{k}}\left(\vec{G}_0\right)
\mathcal{U}^\dagger_{\vec{k}}\left(\vec{G}_0\right)=\mathds{1}$.
The Eq. \pref{gauge_phi} in turn implies:
\bea
\ket{m,\vec{k}+\vec{G}_0}=\sum_n \mathcal{U}_{mn,\vec{k}}\left(\vec{G}_0\right)
\ket{n,\vec{k}}.
\eea
\end{subequations}
In particular, $\mathcal{U}_{\vec{k}}\left(\vec{G}_0\right)=\mathds{1}$ for a periodic gauge.

In the numerical calculations, the number of Fourier components defining the eigenfunctions $\ket{m,\vec{k}}$ is bounded by a cutoff: $|\vec{G}|<G_c$, where $G_c$ is chosen in order to achieve the convergence of the low energy bands.

\section{The continuum model of the commensurate heterostructures: $\text{h}$BN/TBG and $\text{h}$BN/TBG/$\text{h}$BN}
\begin{figure*}
\centering
\includegraphics[scale=0.35]{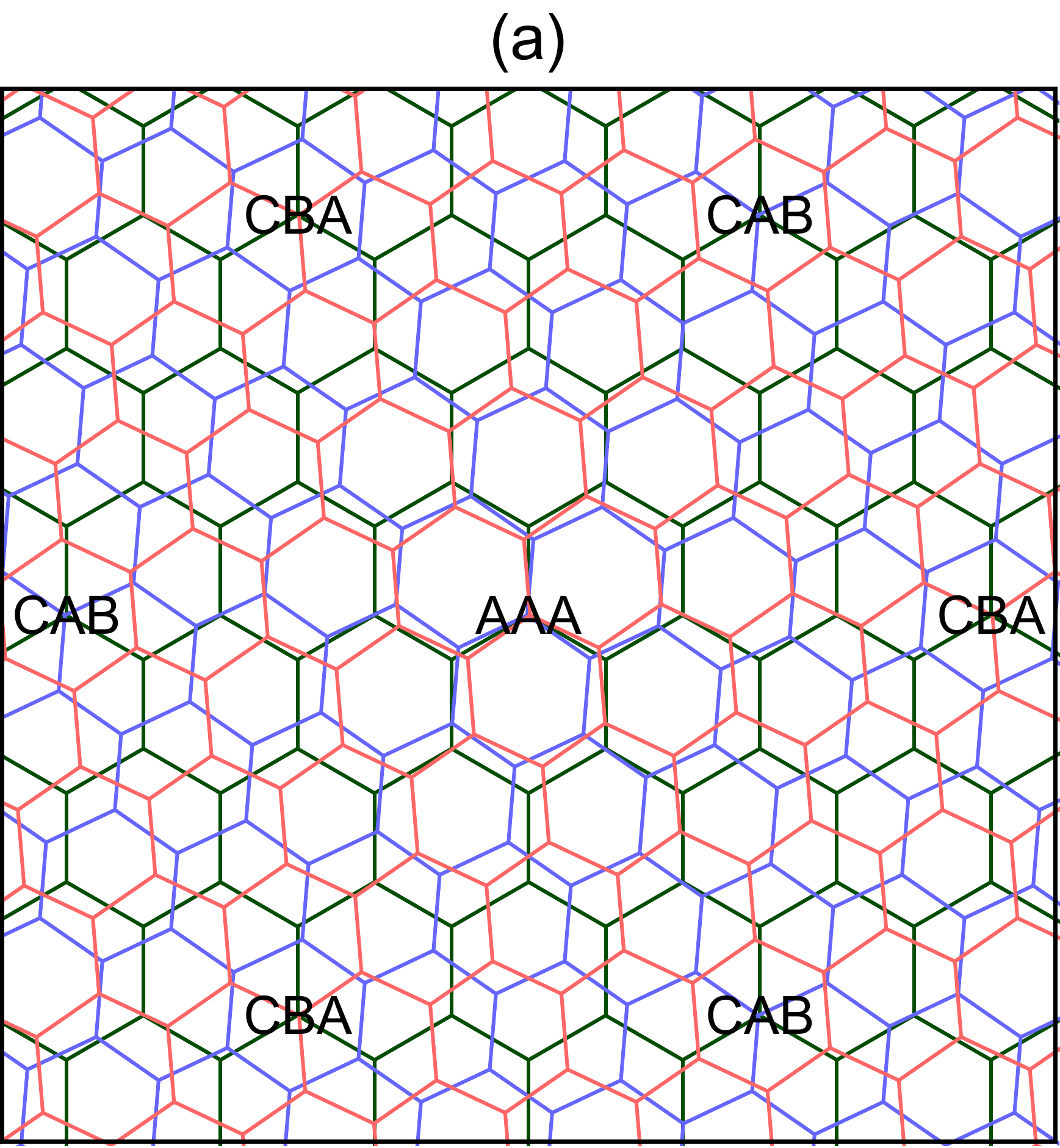}
\includegraphics[scale=0.35]{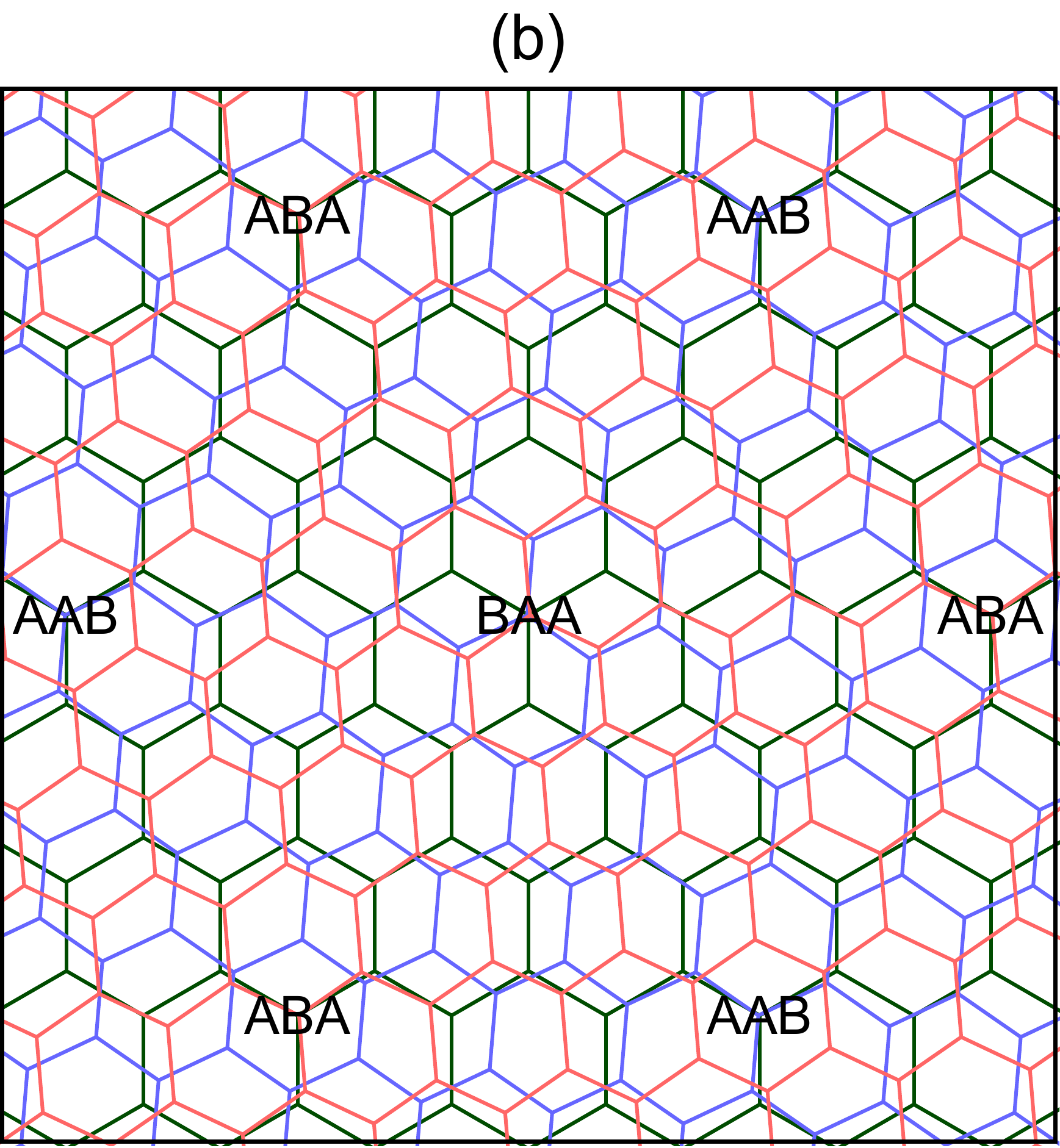}
\includegraphics[scale=0.35]{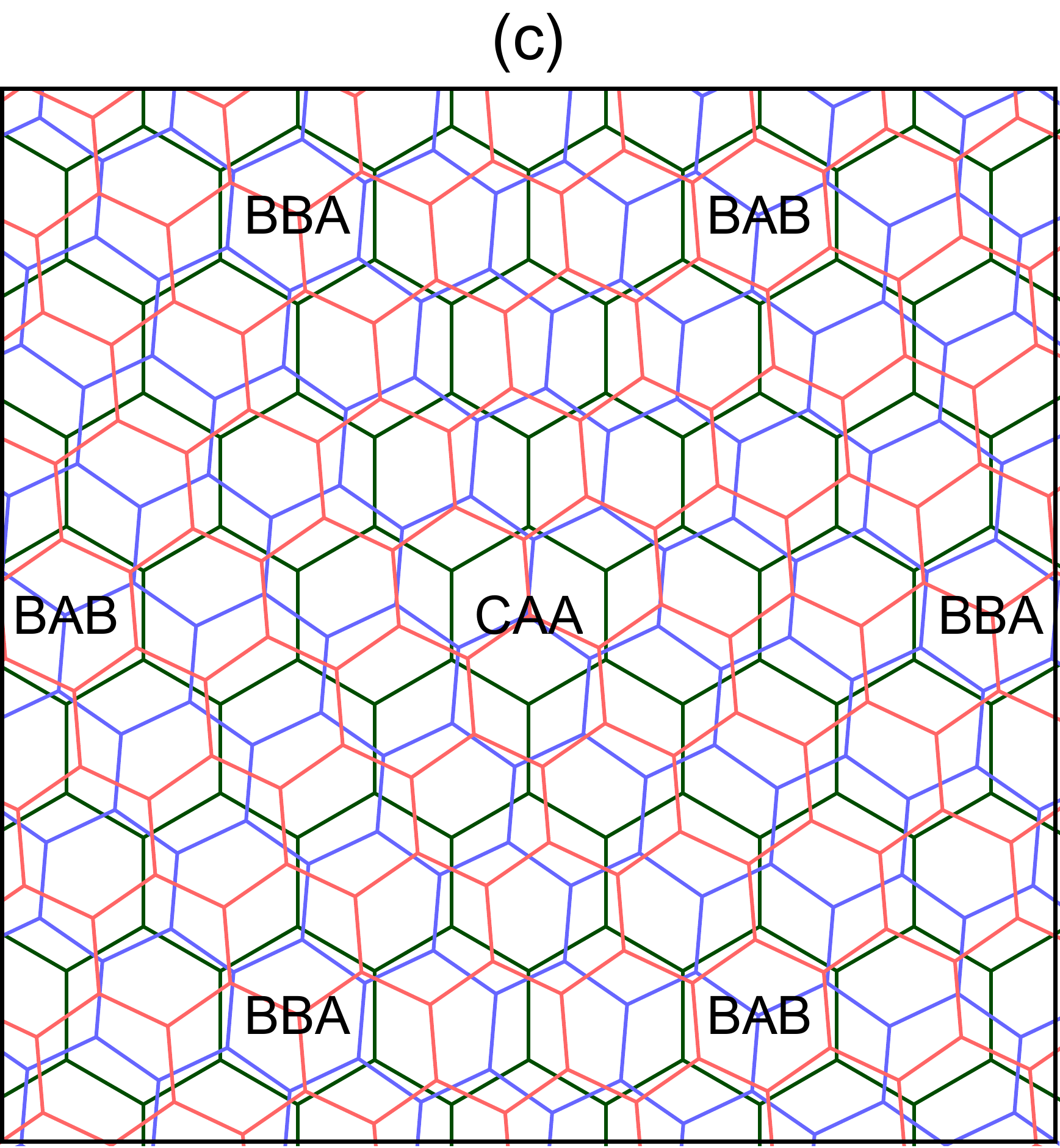}
\caption{
Non-equivalent configurations of the moir\'e unit cell.
Green, light blue and red hexagons refer to the hBN, bottom and top grahene layer, respectively, and $C$ denotes the center of the hexagon.
}
\label{moire_cells}
\end{figure*}
\begin{figure}
\centering
\includegraphics[scale=0.4]{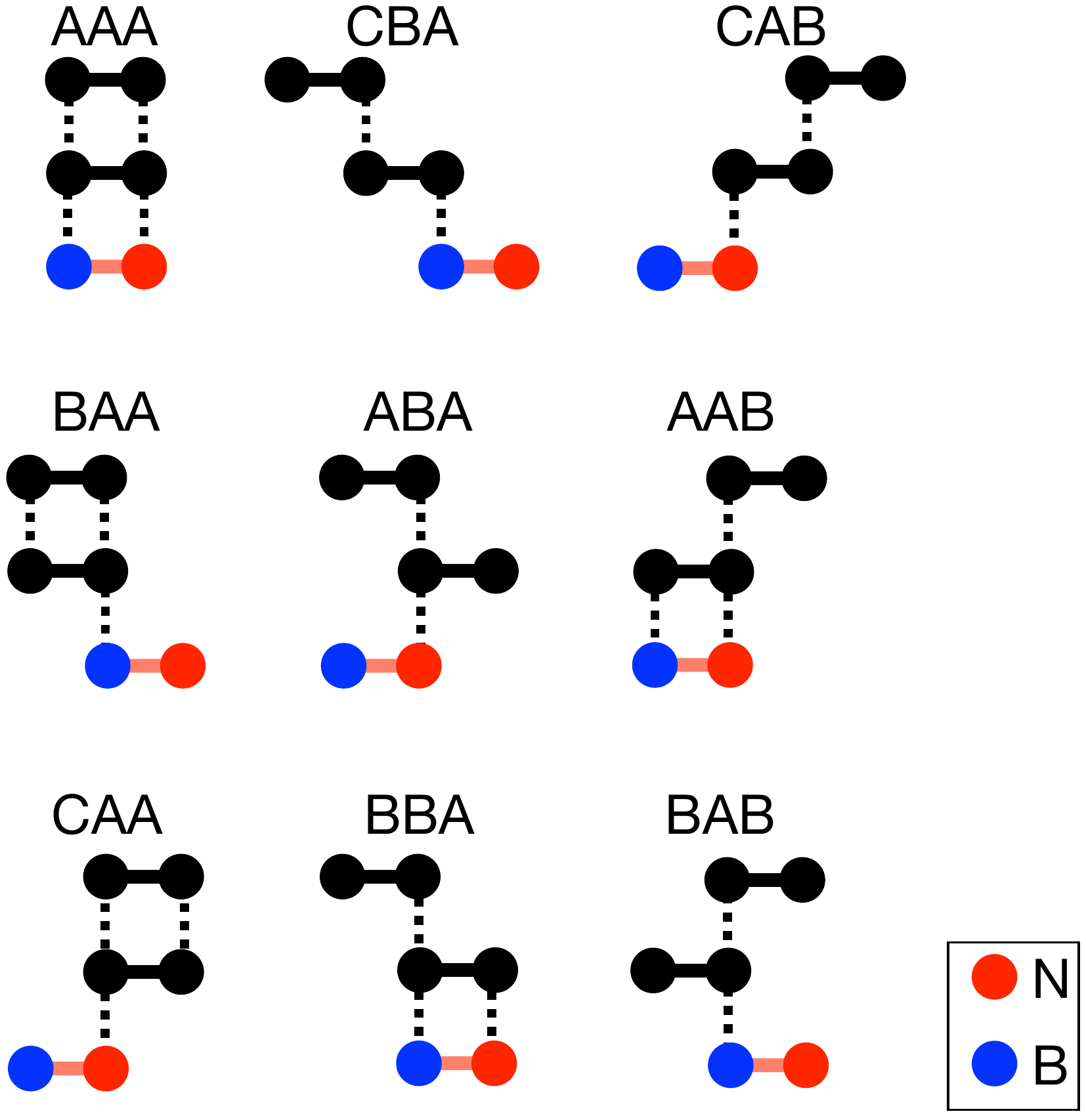}
\caption{
Schematic representation of the nine stacking arrangements appearing in the configurations of the Fig. \ref{moire_cells}.
The black points represent the carbon atoms.
The points at the right (left) edges of the horizontal lines represent atoms of type $A$ ($B$).
We are assuming that the sub-lattice $A$ of the hBN is occupied by nitrogen atoms.
}
\label{stackings}
\end{figure}
The commensurate heterostructure hBN/TBG accounts for three possible non-equivalent configurations of the moir\'e unit cell,
as shown in Fig. \ref{moire_cells}, where the green, light blue and red hexagons refer to the hBN, bottom and top graphene layer, respectively,
and $C$ denotes the center of the hexagon.
Each configuration is determined uniquely by the local stacking between the hBN and the TBG in the center of the cell: $AAA$, $BAA$ and $CAA$.
Consequently, there are nine possible stacking arrangements, three for each configuration. They are shown schematically in Fig. \ref{stackings},
where the black points represent the carbon atoms, the points at the right (left) edges of the horizontal lines represent atoms of type $A$ ($B$),
and we assume that the sub-lattice $A$ of the hBN is occupied by nitrogen atoms, without loss of generality.
The case of the heterostructure hBN/TBG/hBN is more complex, accounting for nine non-equivalent configurations, which are determined by the local stacking between the bottom and top hBN and the TBG                                                  in the center of the cell. 
We describe the effect of hBN on the nearest graphene layer by means of an effective potential, periodic in the moir\'e unit cell~\cite{Wallbank2013}:
\bea
V^\alpha_\text{SL}\left(\vec{r}\right)=
w^\alpha_{0}\tau_{0}+\Delta^\alpha\tau_{z}+
\sum_{j=0}^5
v^\alpha_\text{SL}(\vec{G}_j)e^{i\vec{G}_j\cdot\vec{r}},
\eea
where $\alpha=A,B,C$ labels the three configurations of the Fig. \ref{moire_cells}(a),(b),(c), respectively,
$w^\alpha_{0}$ and $\Delta^\alpha$ represent a spatially uniform scalar and mass term (note that hBN breaks inversion symmetry, and allows for a mass term~\cite{Hunt2013}),
$\vec{G}_j=\frac{4\pi}{\sqrt{3}L}\left(\cos\frac{\pi (j+1)}{3},\sin\frac{\pi (j+1)}{3}\right)$ identify the first star of reciprocal lattice vectors and we assume that the modulation of $V^\alpha_\text{SL}$ at smaller wavelengths is negligible.
The amplitudes $v^\alpha_\text{SL}(\vec{G}_j)$ are given by:
\bea\label{eq: hbN Perturbation}
v^\alpha_\text{SL}(\vec{G}_j)=
\left[V_{s}^{e,\alpha}+i(-1)^{j}V_{s}^{o,\alpha}\right]\tau_{0}
 +  \left[V_{\Delta}^{o,\alpha}+i(-1)^{j}V_{\Delta}^{e,\alpha}\right]\tau_{z}+\left[V_{g}^{o,\alpha}+i(-1)^{j}V_{g}^{e,\alpha}\right]M_j,
\eea
where $M_j=\left(iG_{j}^{x}\tau_{y}-iG_{j}^{y}\tau_{x}\right)/\left|\vec{G}_j\right|$.
The parameters $V_{s}^{e,\alpha}$ and $V_{s}^{o,\alpha}$ are position-dependent scalar terms and are even and odd under spatial inversion, respectively. Similarly, $V_{\Delta}^{o(e),\alpha}$ and $V_{g}^{o(e),\alpha}$ are position-dependent mass and gauge terms, respectively. We use the parametrization of $V^\alpha_{\text{SL}}$ given by the Ref.~\cite{Jung2017}. In particular, the set of parameters for the configuration: $\alpha=A$, is:
\bea\label{vHBN_par}
\left(w^A_{0},\Delta^A,V_{s}^{e,A},V_{s}^{o,A},V_{\Delta}^{e,A},V_{\Delta}^{o,A},V_{g}^{e,A},V_{g}^{o,A}\right)=
\left(0, 3.62, -1.874, 6.775, 0.017, -6.849, 3.609, -12.43\right)\text{meV}.
\eea
The configurations: $\alpha=B,C$, are related to $\alpha=A$ by a rotation of $\pm 2\pi/3$ in the parameters space, as detailed in Ref.~\cite{Jung2017}.  

The continuum Hamiltonians of the hBN/TBG and of the hBN/TBG/hBN are given by:
\begin{subequations}\label{H_heterostructures}
\bea
H^\alpha_\text{hBN/TBG}&=&
\begin{pmatrix}
H_b+V^\alpha_\text{SL}(\vec{r})&U(\vec{r})\\U^\dagger(\vec{r})&H_t
\end{pmatrix},\\
H^{\alpha\beta}_\text{hBN/TBG/hBN}&=&
\begin{pmatrix}
H_b+V^\alpha_\text{SL}(\vec{r})&U(\vec{r})\\U^\dagger(\vec{r})&H_t+V^\beta_\text{SL}(-\vec{r})
\end{pmatrix},
\eea
\end{subequations}
respectively.

\section{Topological phases induced by the Hartree Potential}

In this section, we address the topological phases of the bands for the three possible non-equivalent configurations, where the breaking of either time reversal or inversion symmetry in the single valley model in Eq.~\ref{H_heterostructures} allows for a finite Berry curvature
\begin{equation}
\Omega_{\vec{k},l} = 2~\mathrm{Im} \braket{\partial_{k_x}\psi_{\vec{k},l}|\partial_{k_y}\psi_{\vec{k},l}},
\label{eq:Omega}
\end{equation}
where $l$ is a band index with energy $E_l\left(\vec k\right)$ and wavefunctions $\psi_{\vec{k},l}$. Notice that due to time\textendash reversal symmetry, the Berry curvature in each graphene valley has opposite sign and hence the total Chern number is zero. However, by assuming an absence of intervalley scattering, the topological invariants can be defined separately~\citep{San-Jose2014a,Song2015}. For the different stacking configurations, the bands are isolated and their Berry curvature is well defined. Therefore, we can assign a valley Chern number ${{\cal C}_{l}}$ to the band $l$ which is given by the integral of the Berry curvature about the moir\'e Brillouin zone:
\begin{equation}
\label{eq: Chern}
{\cal C}_{l} = \frac{1}{2 \pi} \int_\mathrm{mBZ} d^2 \vec k \Omega_{\vec{k},l}.
\end{equation}
We use the algorithm in Ref.~\cite{Fukui2005} to compute the Berry curvature and valley Chern number. The presence of an hBN substrate gives rise to an insulating structure  with  a  small  band  gap,  which  is  due  to  the breaking  of  inversion  symmetry  due  to  the  hBN  layer. As is shown in  Fig. \ref{Fig: BandAA} in the main text, the resulting isolated conduction and valence bands in each valley and in each stack configuration carry Chern numbers ${\cal C}=\pm 1$, in agreement with Ref.~\cite{Zhang2019b,bultinck_cm2019}. However, similar to other graphene heterostructures \citep{PCBWG20} and consistent with experimental observations \cite{Serlin2019, Xiaobo_nat19}, we find different topological phases induced by interaction effects.  This is summarized in Table.~\ref{tab: TablaChern} where we display the Chern number as a function of the filling of the conduction band, for the different stack configurations.

\begin{table}
\begin{centering}
\begin{tabular}{|c|c|c||c|c||c|c|}
\cline{2-7} \cline{3-7} \cline{4-7} \cline{5-7} \cline{6-7} \cline{7-7} 
\multicolumn{1}{c|}{} & \multicolumn{2}{c|}{AAA} & \multicolumn{2}{c|}{CAA} & \multicolumn{2}{c|}{BAA}\tabularnewline
\hline 
$\nu$ & ${\cal C}_{b}$ & ${\cal C}_{t}$ & ${\cal C}_{b}$ & ${\cal C}_{t}$ & ${\cal C}_{b}$ & $C_{t}$\tabularnewline
\hline 
$-4$ & $2$ & $-2$ & $-2$ & $3$ & $-2$ & $2$\tabularnewline
\hline 
$-3$ & $2$ & $-2$ & $1$ & $0$ & $-2$ & $2$\tabularnewline
\hline 
$-2$ & $2$ & $-2$ & $1$ & $0$ & $1$ & $-1$\tabularnewline
\hline 
$-1$ & $-1$ & $1$ & $1$ & $0$ & $1$ & $-1$\tabularnewline
\hline 
$0$ & $-1$ & $1$ & $1$ & $-1$ & $1$ & $-1$\tabularnewline
\hline 
$1$ & $-1$ & $1$ & $1$ & $-1$ & $1$ & $-1$\tabularnewline
\hline 
$2$ & $-1$ & $1$ & $1$ & $-1$ & $1$ & $-1$\tabularnewline
\hline 
$3$ & $-1$ & $1$ & $1$ & $-1$ & $1$ & $-1$\tabularnewline
\hline 
$4$ & $-1$ & $1$ & $1$ & $-1$ & $1$ & $-1$\tabularnewline
\hline 
\end{tabular}
\par\end{centering}
\caption{Valley Chern number as a function of the filling of the conduction
band for the different stack configurations. ${\cal C}_{b}$ and ${\cal C}_{t}$ are the Chern numbers of the bottom and top narrow bands of Fig. \ref{Fig: BandAA} in the main text.}
\label{tab: TablaChern}
\end{table}

\section{Changes in the charge density induced by the substrate}
\begin{figure*}
    \centering
    \includegraphics[scale=0.35]{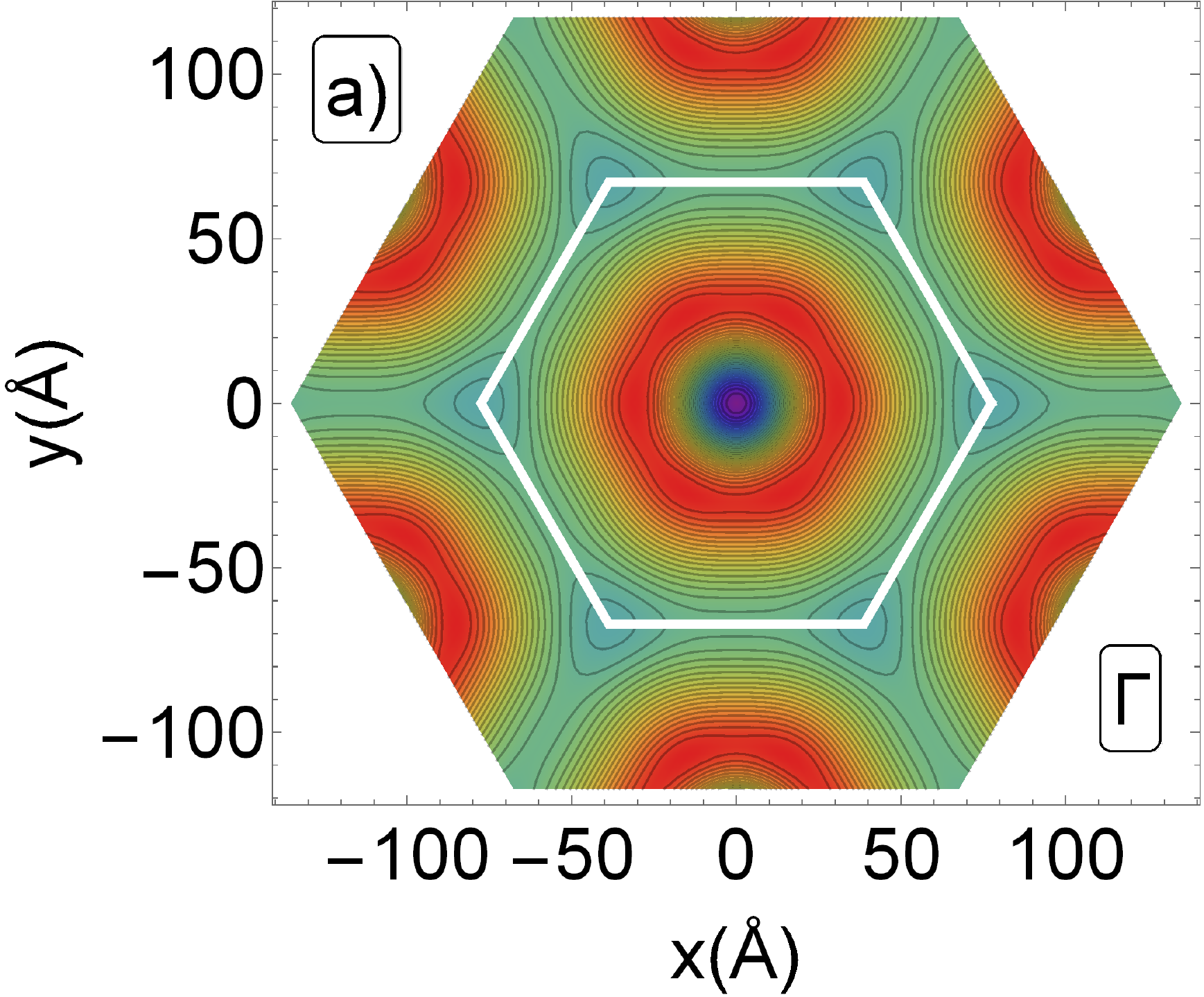}
    \includegraphics[scale=0.35]{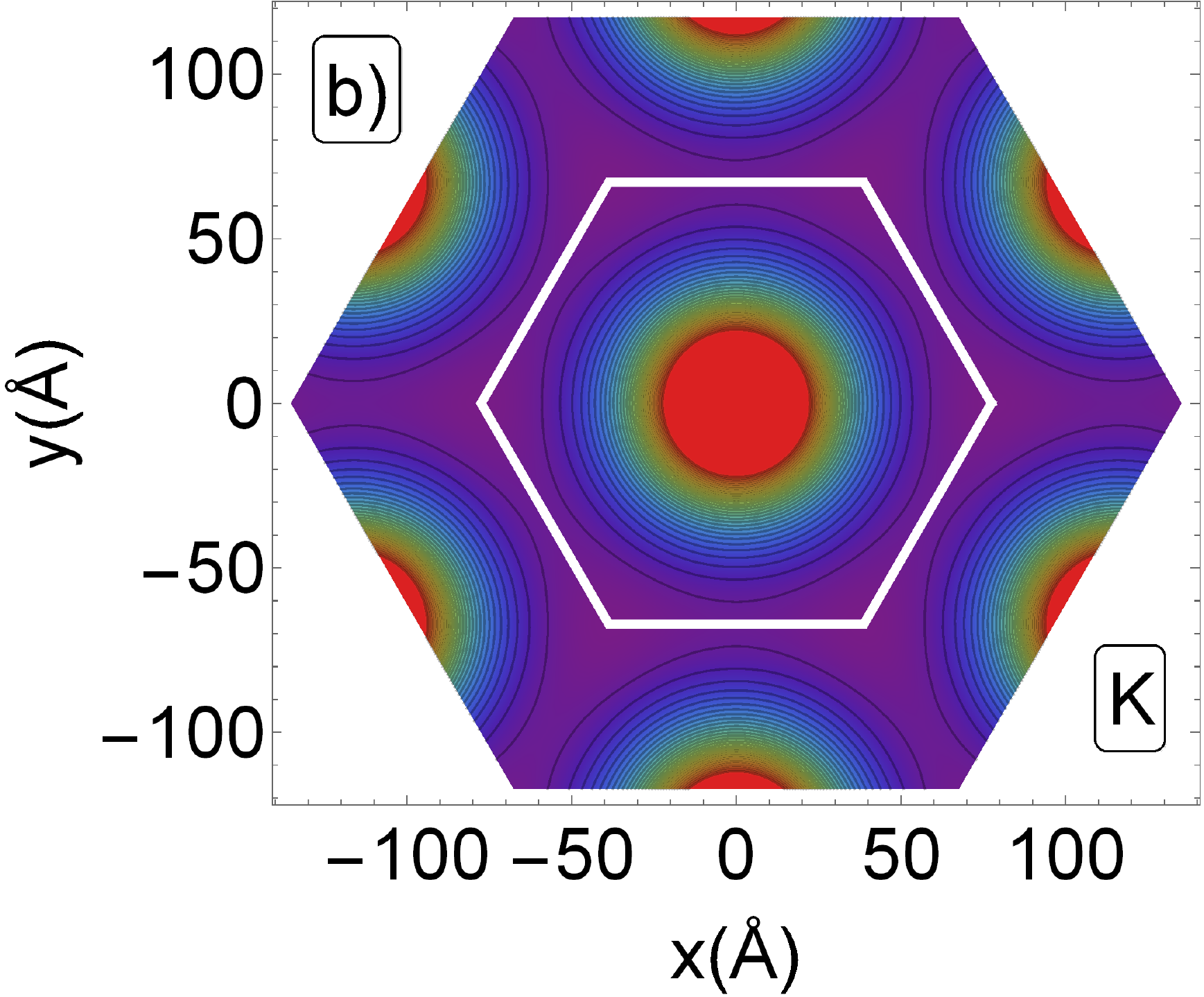} \\
    \includegraphics[scale=0.35]{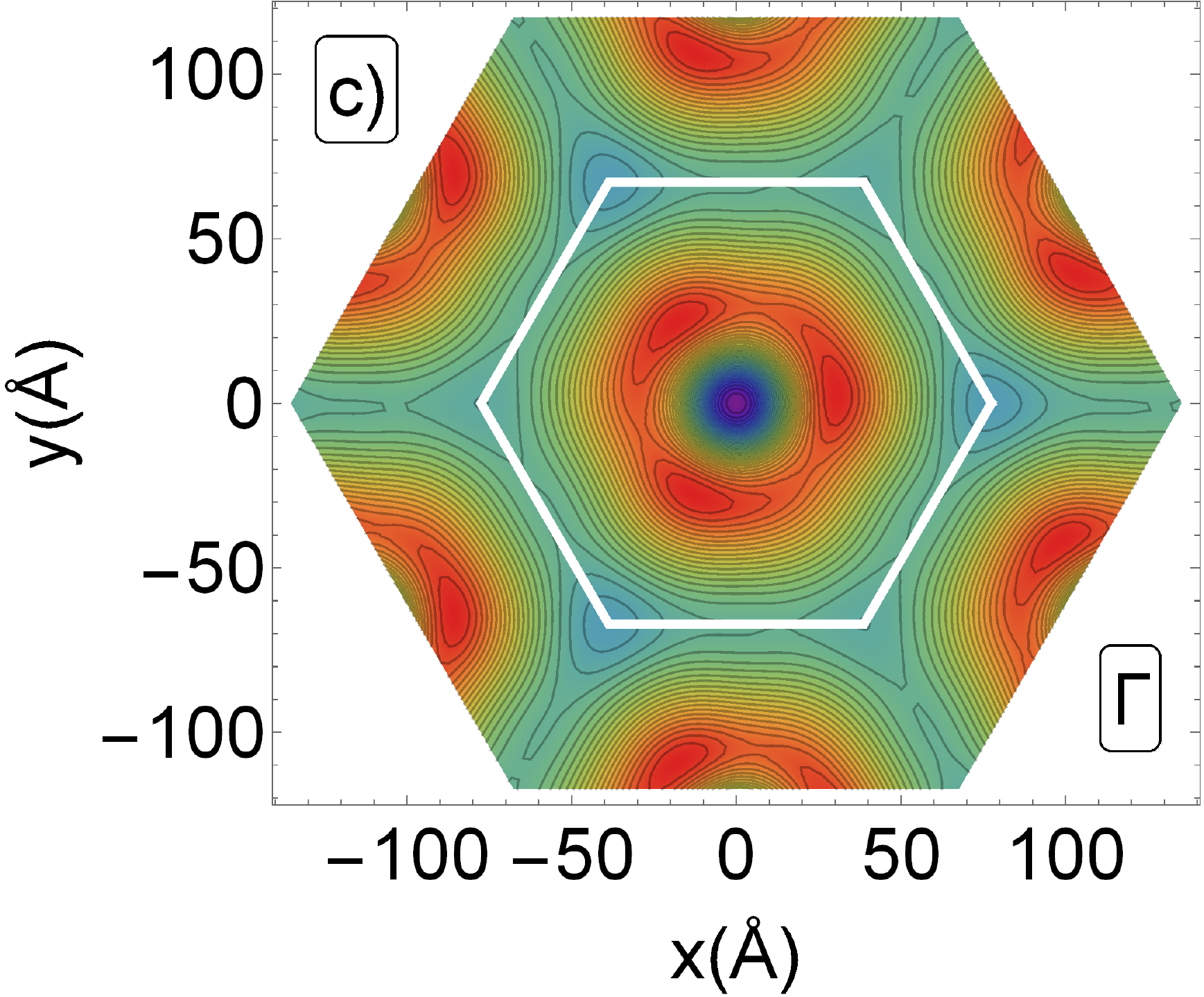}
    \includegraphics[scale=0.35]{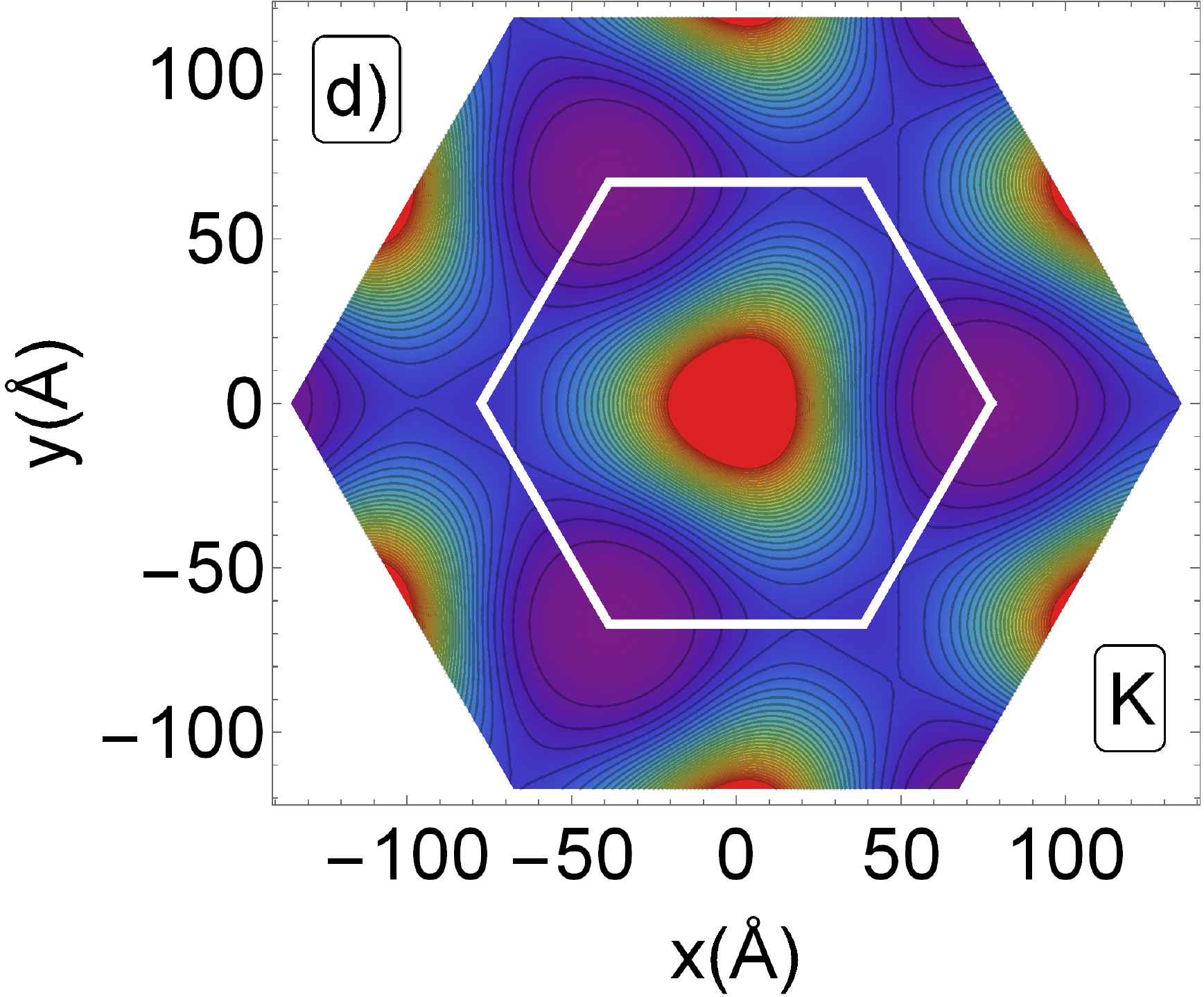}
    \caption{Charge densities of states for the pristine TBG (top row) and TBG on top of hBN (bottom row) for the AAA stack configuration. The white hexagon is the real-space unit cell. The scale is arbitrary, with purple denoting the minimum of the charge density, and red the maximum.} 
    \label{charge density}
\end{figure*}

In Fig. \ref{charge density} we show the real space charge density computed at the points $\Gamma$ and $K$ of the BZ, and obtained for pristine TBG, panels a) and b), and for TBG with a substrate of hBN, panels c) and d). Interestingly, we find that the $\mathcal{C}_6$ symmetry of the charge density of pristine TBG is lowered to $\mathcal{C}_3$ by the presence of the substrate, an effect that is more evident at the $K$ point.   

\section{Hartree-Fock band structure of the commensurate hBN/TBG}
\begin{figure*}
    \centering
    \includegraphics[scale=0.35]{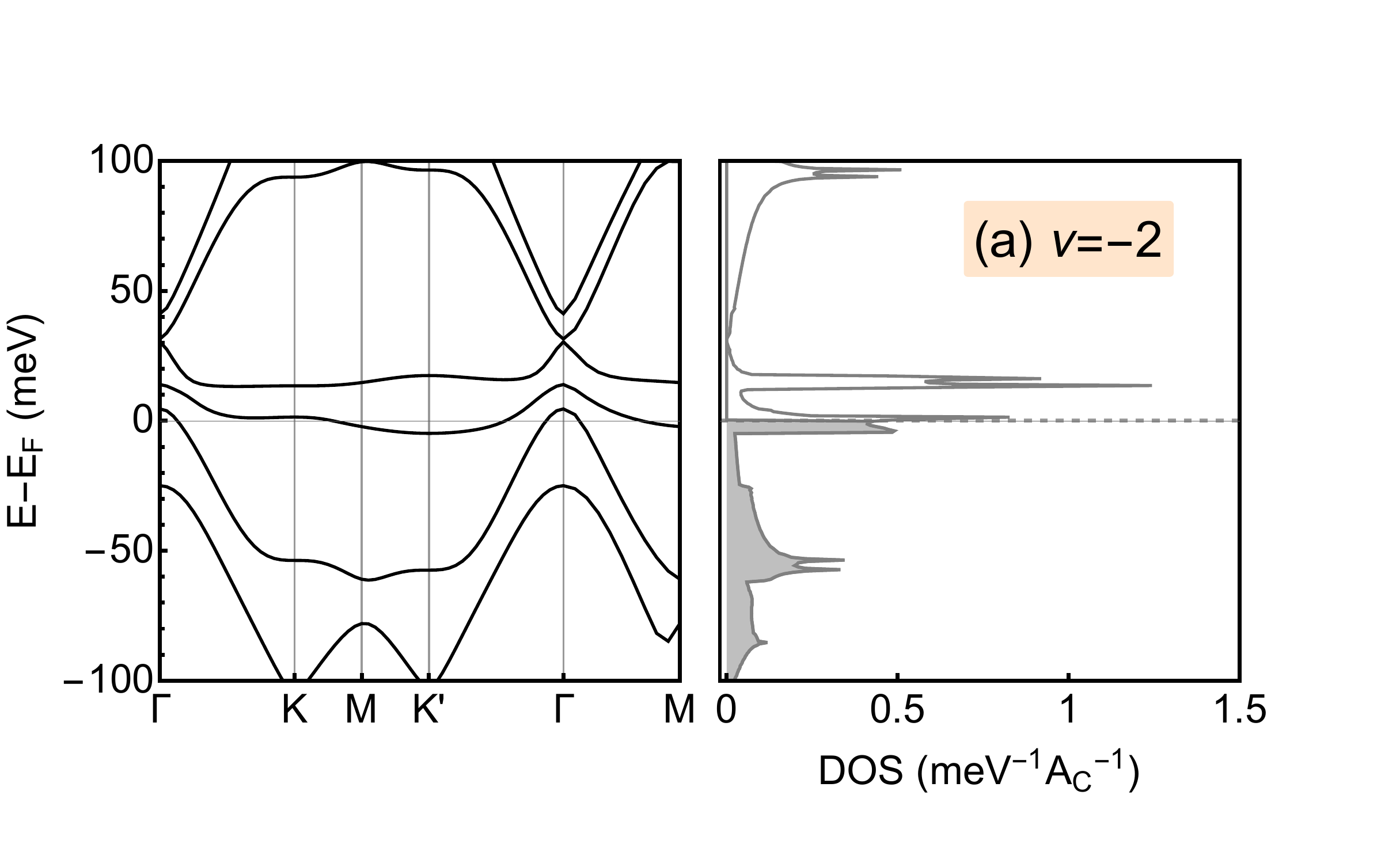}
    \includegraphics[scale=0.35]{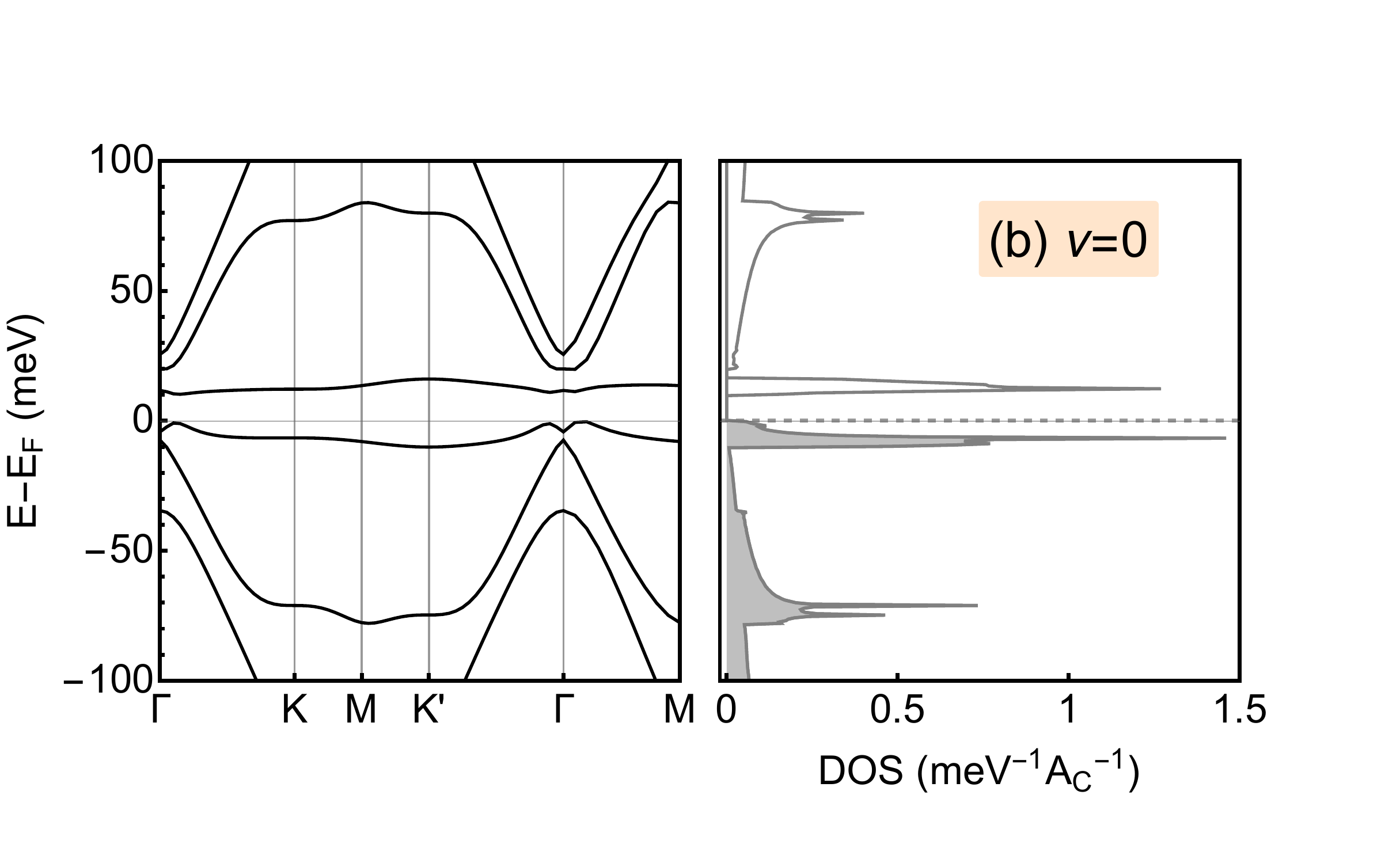}\\
    \includegraphics[scale=0.35]{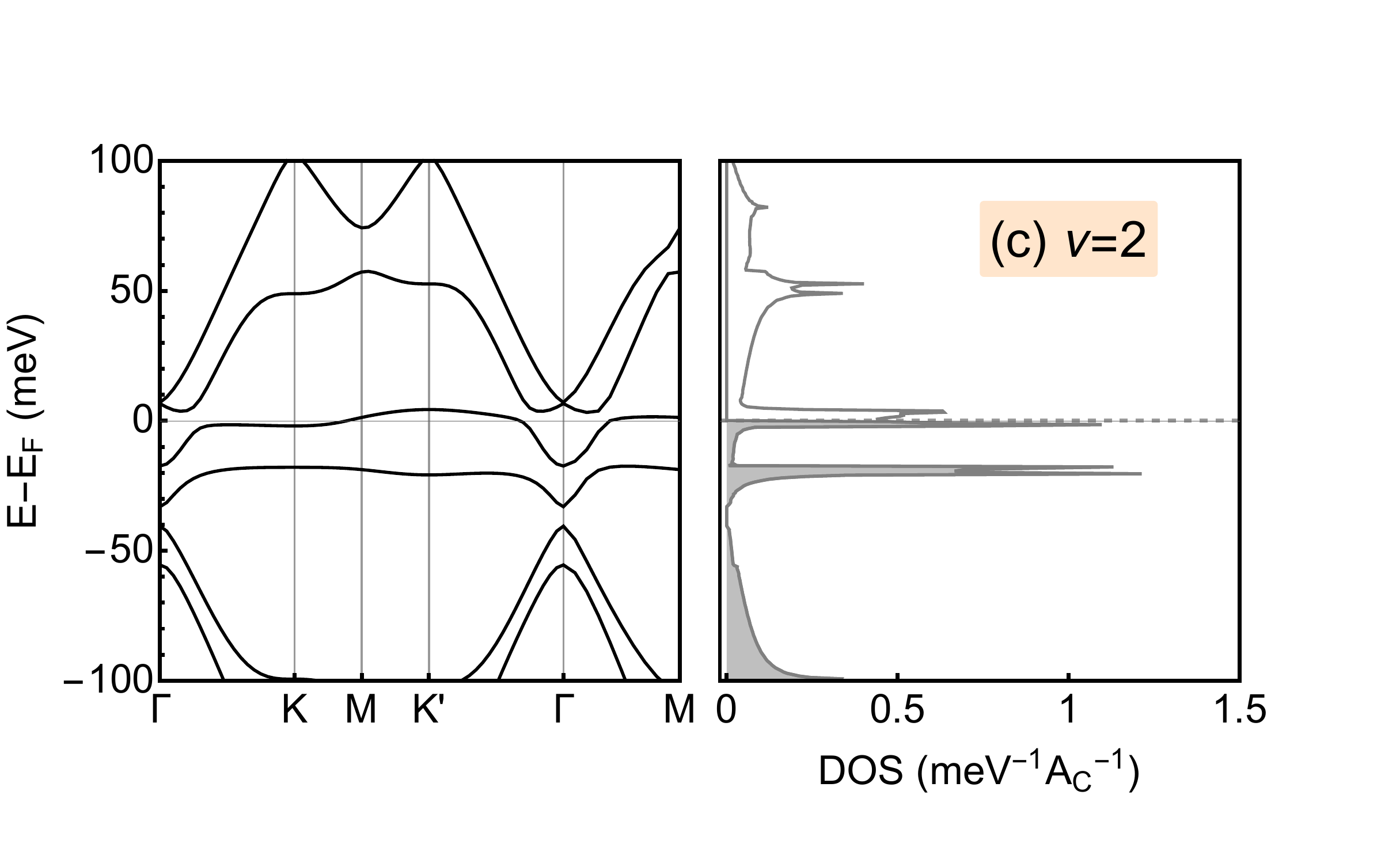}
    \includegraphics[scale=0.35]{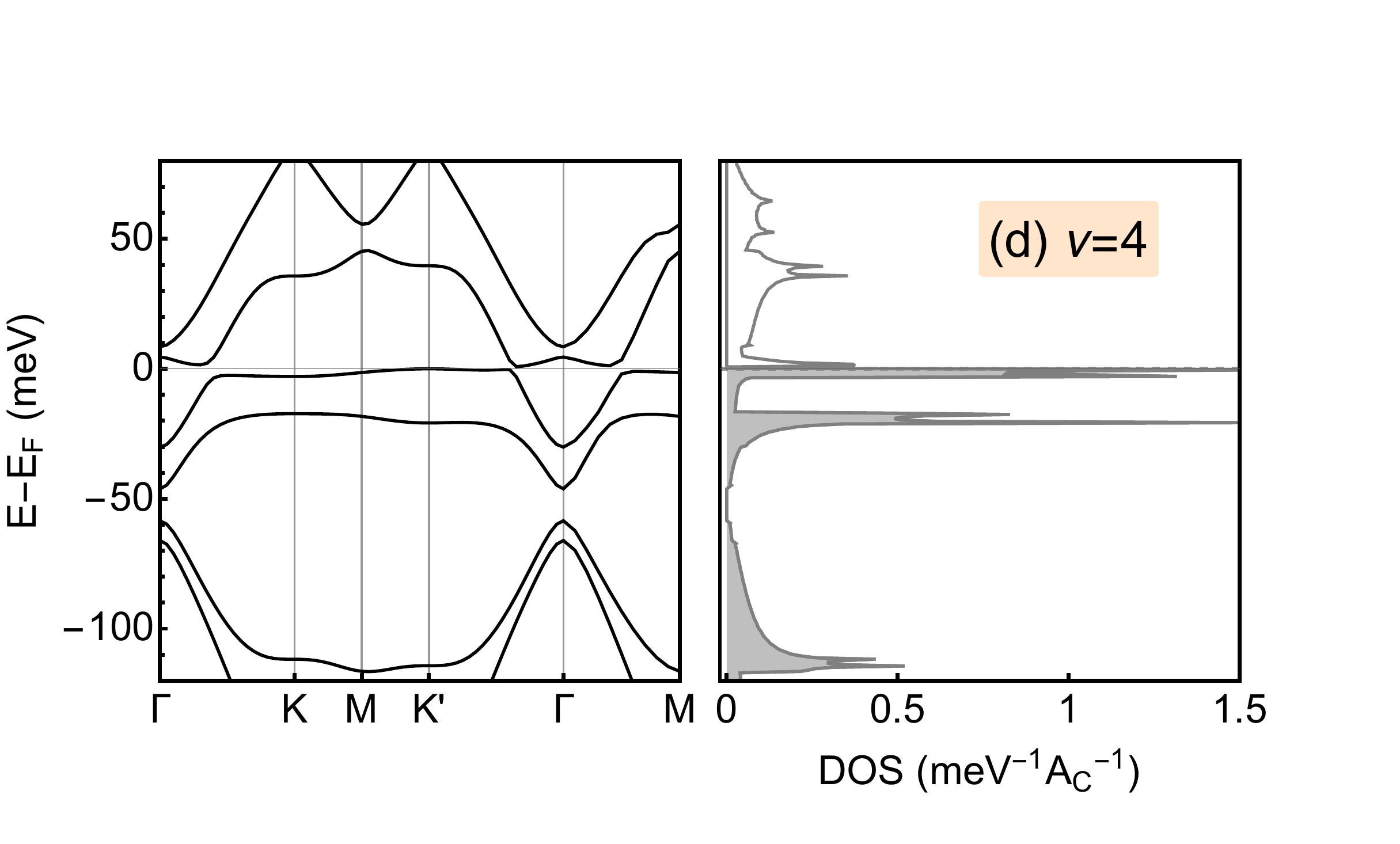}
    \caption{Band structure and DOS of the commensurate heterostructure hBN/TBG, obtained within the full Hartree-Fock approximation, for different fillings of the conduction band, $\nu$.}
    \label{Hartree_Fock_bands}
\end{figure*}

Using the formalism detailed in\cite{CG20},
we performed a fully Hartree-Fock calculation of the band structure of the commensurate hBN/TBG heterostructure, in the stacking configuration $\alpha=A$.
The results are shown in the Fig. \ref{Hartree_Fock_bands} for different fillings of the conduction band, $\nu$.
As it is evident, the spectra display pinning of the Fermi level, $E_F$, at the van Hove singularity at finite filling and are quite similar to the ones shown in the main text, which include only the Hartree potential.
Consequently, we argue that the Fock contribution is negligible for this kind of system, in contrast to the case of freely standing TBG, where the Fock terms have been shown to change considerably the band structure and to open spectral gaps at integer filings\cite{Xie2020,ZJWZ20,CG20}. 

\section{The case of non-commensurate heterostructures of $\text{h}$BN/TBG}
Here we focus on the general case in which the moir\'e pattern identified by the hBN and its nearest graphene layer and that
of the TBG are not commensurate, making the system non-periodic. This happens, for example, when the hBN is perfectly aligned with its nearest graphene layer.

We compute the quasi-band structure and the DOS, shown in the Fig. 5 of the main text, by projecting the perturbation induced by the hBN on the low energy eigenstates of the TBG.

Let us write the local potential induced by the hBN on the nearest graphene layer as:
\bea
V_{\text{SL}}(\vec{r})=V_0+\delta V_{\text{SL}}(\vec{r}),
\eea
where $V_0=w_0\tau_0+\Delta\tau_z$ is the uniform contribution, while
$$\delta V_{\text{SL}}(\vec{r})=\sum_{j=0}^5 v_{\text{SL}}\left(\vec{\tilde{G}}_j\right)e^{i\vec{\tilde{G}}_j\cdot\vec{r}}$$
is the contribution at finite wavelengths. In the following we choose the parametrization of $V_{\text{SL}}$ given by the Eq. \pref{vHBN_par}, which corresponds to set the origin in a region of AAA stacking.
The $\vec{\tilde{G}}_j$ are reciprocal vectors of the moir\'e superlattice identified by the hBN/G, as given by:
\bea\label{Gtilde}
\vec{\tilde{G}}_0=
R\left(-\theta/2\right)\left[
\mathds{1}-(1+\delta)^{-1}R\left( \theta_{hBN} \right)
\right]
\begin{pmatrix}
0\\1
\end{pmatrix}\times
\frac{4\pi}{\sqrt{3}d_G}
\quad,\quad
\vec{\tilde{G}}_j=R\left(\pi j/3\right)\vec{\tilde{G}}_0,
\eea
where $R$ is the rotation matrix in two dimensions and $\delta=d_{hBN}/d_G-1\simeq0.017$ is the lattice mismatch between hBN and graphene.
The overall rotation of $-\theta/2$ in the lhs of the Eq. \pref{Gtilde} takes into account the absolute orientation of the bottom graphene layer of the TBG,
which we are assuming to be the closest one to the hBN.
Note that $\vec{\tilde{G}}_j\ne \vec{G}$ for any reciprocal vector of the moir\'e superlattice of the TBG, $\vec{G}$, as long as the two moires are not commensurate. Consequently, the full Hamiltonian including both $H_{\text{TBG}}$ and $V_{\text{SL}}$ cannot be diagonalized by any Bloch wave.

We proceed by first diagonalizing the Hamiltonian $H_{\text{TBG}}$ in the presence of the uniform term, $V_0$, which does not modify the periodicity of the TBG.
The resulting eigenfunctions are Bloch waves, $\ket{m,\vec{k}}$, as expressed by the Eq. \pref{Bloch_waves}.
The Hamiltonian of the hBN/TBG can be generally written in this basis as:
\bea\label{Hfull_proj}
H_{\text{hBN/TBG}}=\sum_{m\vec{k}}E_m(\vec{k})\ket{m,\vec{k}}\bra{m,\vec{k}}+
\sum_{m\vec{k}}\sum_{n\vec{k}'}
\ket{m,\vec{k}}
t_{mn}\left(\vec{k},\vec{k}'\right)
\bra{n,\vec{k}'},
\eea
where $\vec{k},\vec{k}'$ run in the first BZ of the TBG and
$t_{mn}\left(\vec{k},\vec{k}'\right)$ are the matrix elements of $\delta V_{\text{SL}}$, given by:
\bea
t_{mn}\left(\vec{k},\vec{k}'\right)\equiv
\bra{m,\vec{k}}\delta V_{\text{SL}}\ket{n,\vec{k}'}=
\sum_{\vec{G}\vec{G}'}\sum_{j=0}^5\sum_{ab}
\phi^*_{m,\vec{k},a}\left(\vec{G}\right)v^{ab}_{\text{SL}}\left(\vec{\tilde{G}}_j\right)
\phi_{n,\vec{k}',b}\left(\vec{G}'\right)
\delta_{\vec{k}'+\vec{G}',\vec{k}+\vec{G}-\vec{\tilde{G}}_j},
\eea
where the sum over $a,b$ is restricted to the sub-lattice indices of the graphene layer closest to the hBN.
Given $\vec{k},\vec{G}$ and $j$, it exists only one reciprocal lattice vector of the TBG, $\vec{G}_0$, such that the vector:
$\vec{k}+\vec{G}-\vec{\tilde{G}}_j-\vec{G}_0$
belongs to the first BZ. Then we can write:
\bea
\sum_{n\vec{k}'\vec{G}'}
\phi_{n,\vec{k}',b}\left(\vec{G}'\right)
\delta_{\vec{k}'+\vec{G}',\vec{k}+\vec{G}-\vec{\tilde{G}}_j}
\bra{n,\vec{k}'}
=
\sum_n\phi_{n,\vec{k}-\vec{\tilde{G}}_j+\vec{G}-\vec{G}_0,b}\left(\vec{G}_0\right)
\bra{n,\vec{k}-\vec{\tilde{G}}_j+\vec{G}-\vec{G}_0}&=&\nn\\
=
\sum_{nn'n''}
\mathcal{U}_{nn',\vec{k}-\vec{\tilde{G}}_j}\left(\vec{G}-\vec{G}_0\right)
\mathcal{U}^*_{nn'',\vec{k}-\vec{\tilde{G}}_j}\left(\vec{G}-\vec{G}_0\right)
\phi_{n',\vec{k}-\vec{\tilde{G}}_j,b}\left(\vec{G}\right)
\bra{n'',\vec{k}-\vec{\tilde{G}}_j}&=&\nn\\
=\sum_n
\phi_{n,\vec{k}-\vec{\tilde{G}}_j,b}\left(\vec{G}\right)
\bra{n,\vec{k}-\vec{\tilde{G}}_j}&,&
\eea
where we used the Eq.s \pref{gauge_eqs}.
Thus, the Eq. \pref{Hfull_proj} finally becomes:
\bea\label{dual_tb}
H_{\text{hBN/TBG}}=\sum_{mn\vec{k}}
\left[
\delta_{mn}E_m(\vec{k})\ket{m,\vec{k}}\bra{m,\vec{k}}+
\sum_{j=0}^5
\ket{m,\vec{k}}
 t_{mn}\left(\vec{k},\vec{k}-\vec{\tilde{G}}_j\right)
\bra{n,\vec{k}-\vec{\tilde{G}}_j}\right],
\eea
which defines a dual tight binding multi-orbital model in the reciprocal space, where the $\vec{k}$ points act as sites, $E_m(\vec{k})$ are the onsite energies and only the overlaps between nearest neighbor sites in the triangular lattice are allowed. The corresponding hopping integrals are:
\bea
 t_{mn}\left(\vec{k},\vec{k}-\vec{\tilde{G}}_j\right)=
 \sum_{\vec{G}}\sum_{ab}\phi^*_{m,\vec{k},a}\left(\vec{G}\right)
 v^{ab}_{\text{SL}}\left(\vec{\tilde{G}}_j\right)
 \phi_{n,\vec{k}-\vec{\tilde{G}}_j,b}\left(\vec{G}\right).
\eea
 Because the two moires are not commensurate, the above overlaps span ergodically the BZ of the TBG. Note that this dual tight binding framework is similar to that considered in the Ref.\cite{MKS19} for studying the quasi-crystalline electronic structure of the 30$^\circ$-TBG.
 
\begin{figure}
     \centering
     \includegraphics[scale=0.50]{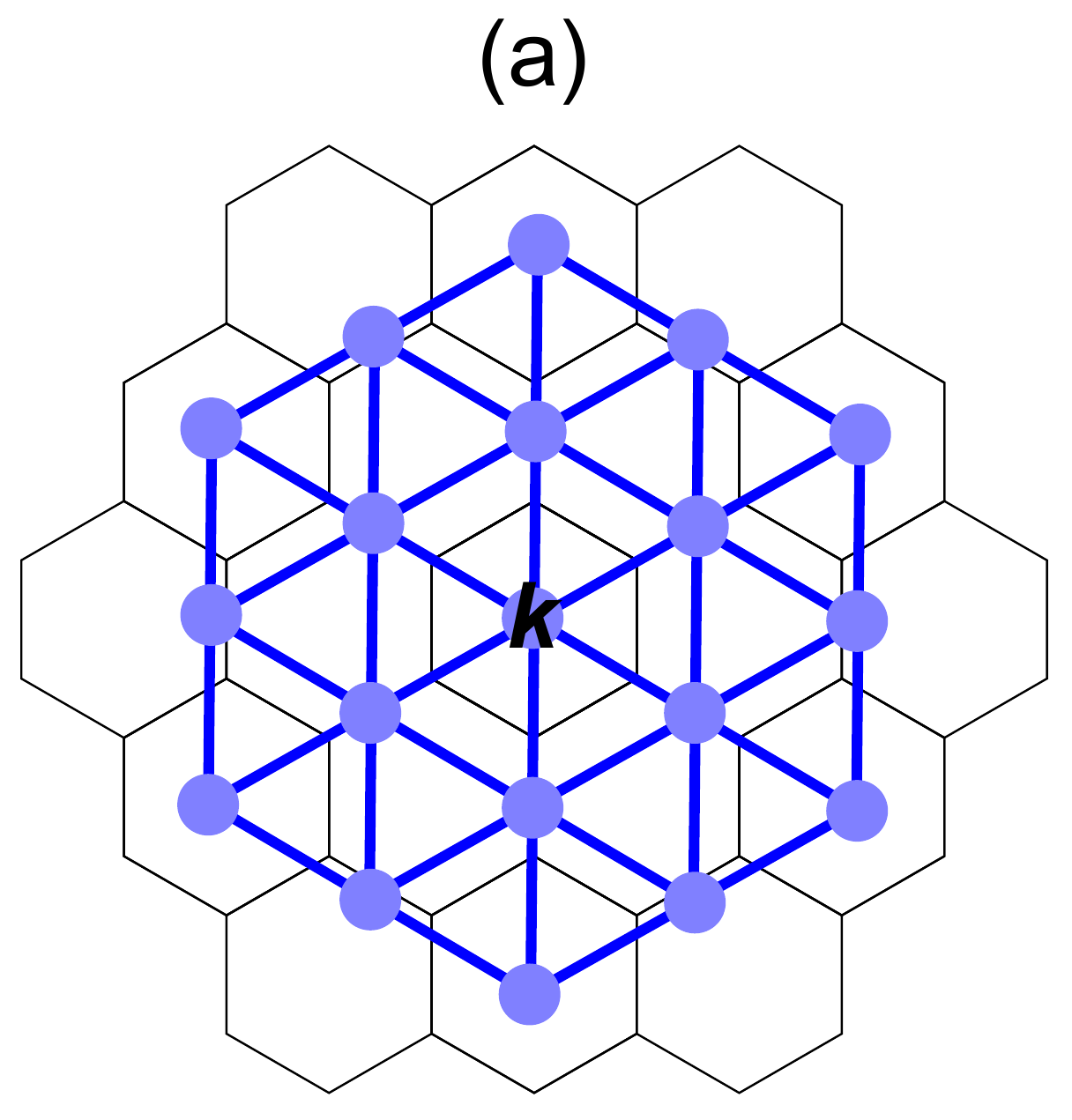}
     \includegraphics[scale=0.50]{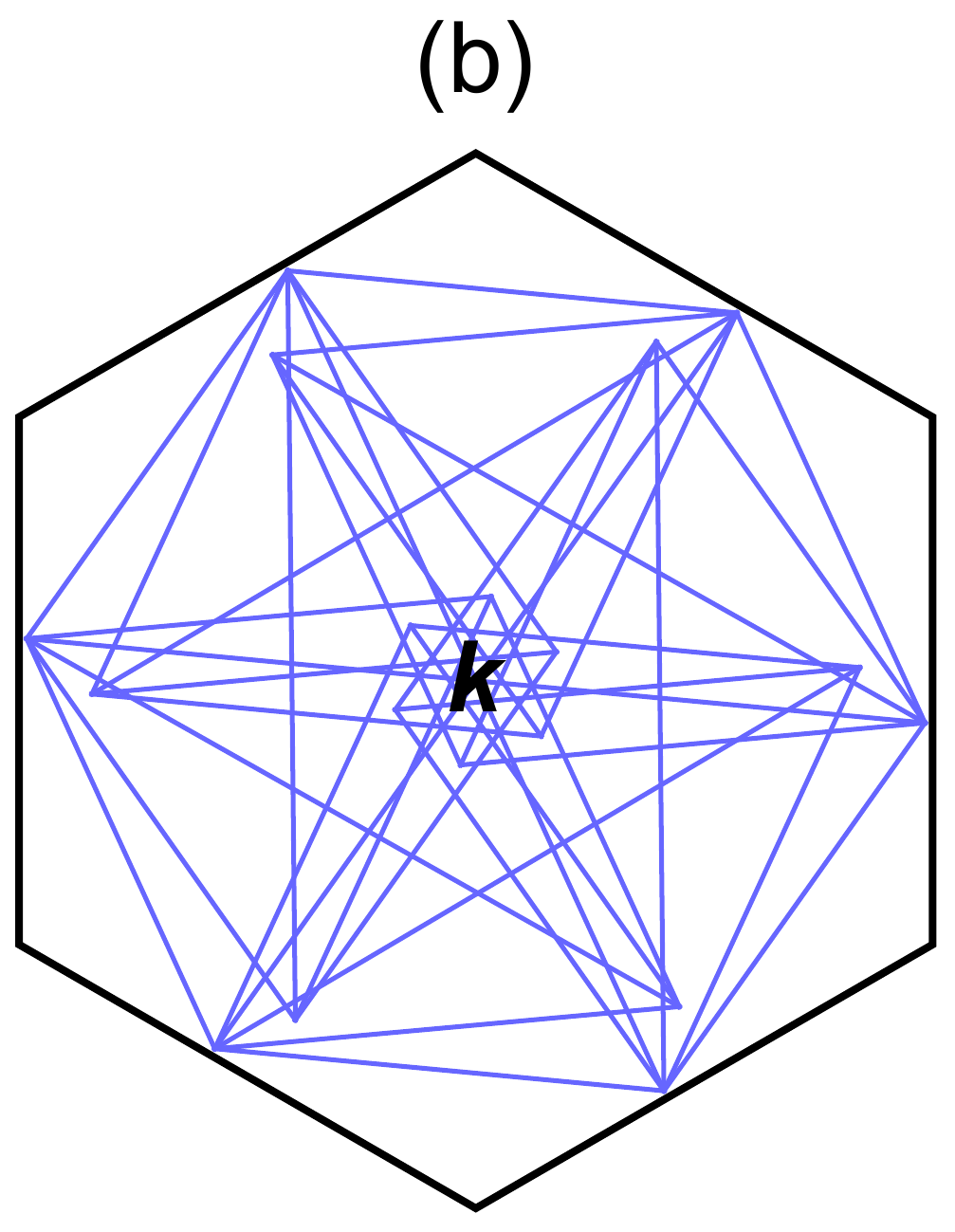}
     \caption{(a) Schematic representation of the dual tight binding model for $\theta_{hBN}=0^\circ$. For each $\vec{k}$ we consider 19 sites (blue points), including $\vec{k}$ itself, and 42 overlaps (blue lines).
     The black hexagons represent the equivalent BZs of the TBG.
     (b) Projection of the overlaps in the first BZ of the TBG.}
     \label{overlaps}
 \end{figure}

 In our calculations, for each $\vec{k}$ we consider the reduced dual model obtained by including only the 19 sites lying in the first two stars of $\vec{\tilde{G}}$ vectors surrounding $\vec{k}$, accounting for 42 overlaps. This is schematically shown in the Fig. \ref{overlaps}(a), where the blue points represent the sites, the blue lines the overlaps between them, the black hexagons are the equivalent BZs of the TBG and $\theta_{hBN}=0^\circ$.
 The Fig. \ref{overlaps}(b) shows the projection of the overlaps in the first BZ of the TBG. Note that, in the commensurate case, all the neighboring sites would collapse into the same point, $\vec{k}$, upon projection.
 We checked that this approximation is sufficient to achieve the convergence of the DOS. Within this framework, the spectrum at the wave vector $\vec{k}$ is obtained by diagonalizing a matrix of size $(19N_b)\times(19N_b)$, where $N_b$ is the number of bands of the TBG taken into account in the projected Hamiltonian of the Eq. \pref{dual_tb}. We consider the $N_b=14$ bands closest to the CN point. The multiple sets of bands shown in the left panels of the Fig. 5 of the main text arise from the diagonalization of the dual tight binding model of the Eq. \pref{dual_tb}. Including more dual sites, there would appear many additional bands which, however, would be just replicas of the ones already shown, but shifted by a different origin in the BZ. As a consequence, they would not change appreciably the DOS and the other physical quantities, as discussed in the Ref.\cite{MKS19}.
 
\begin{figure*}
\centering
\includegraphics[scale=0.3]{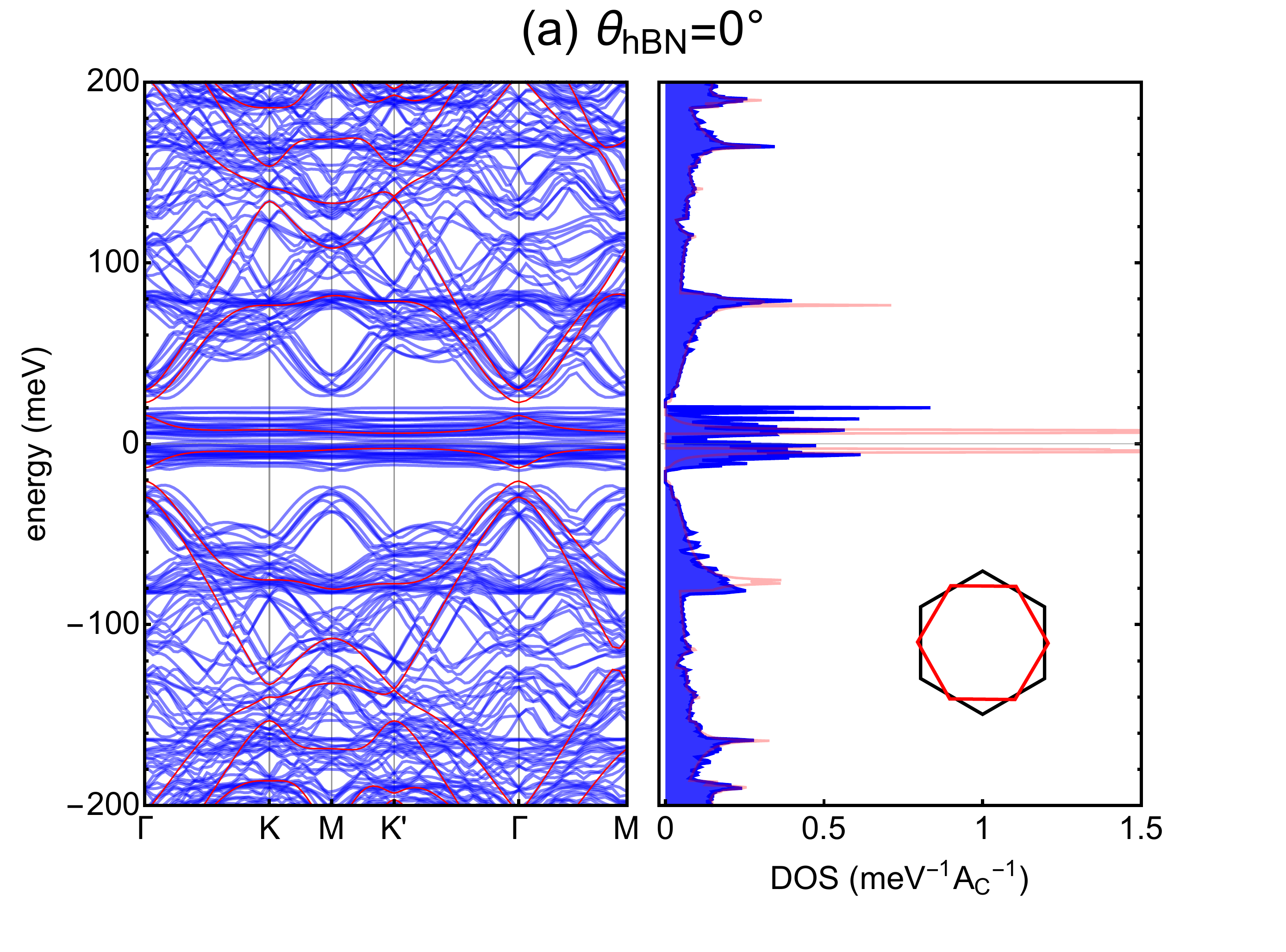}
\includegraphics[scale=0.3]{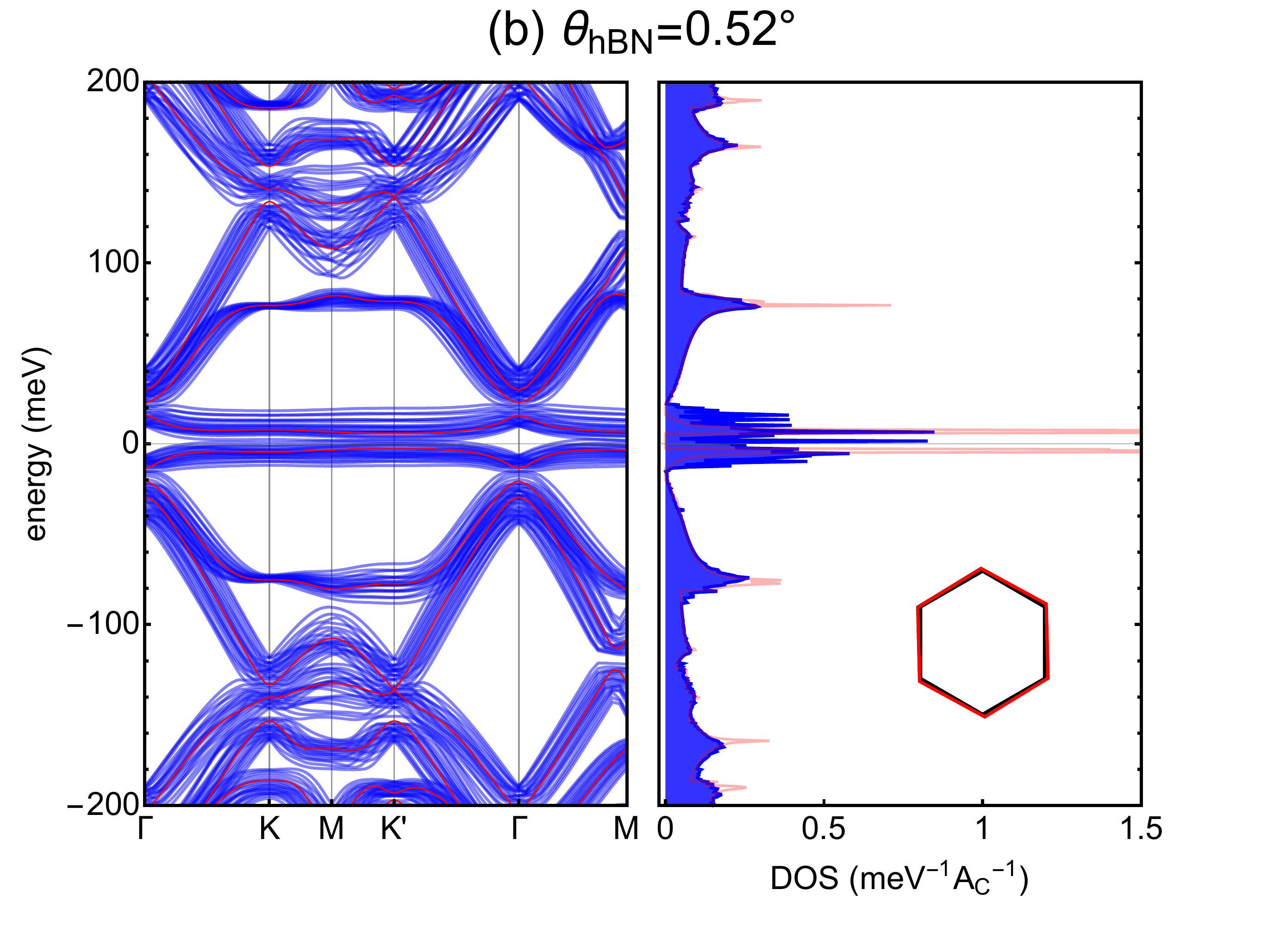}\\
\includegraphics[scale=0.3]{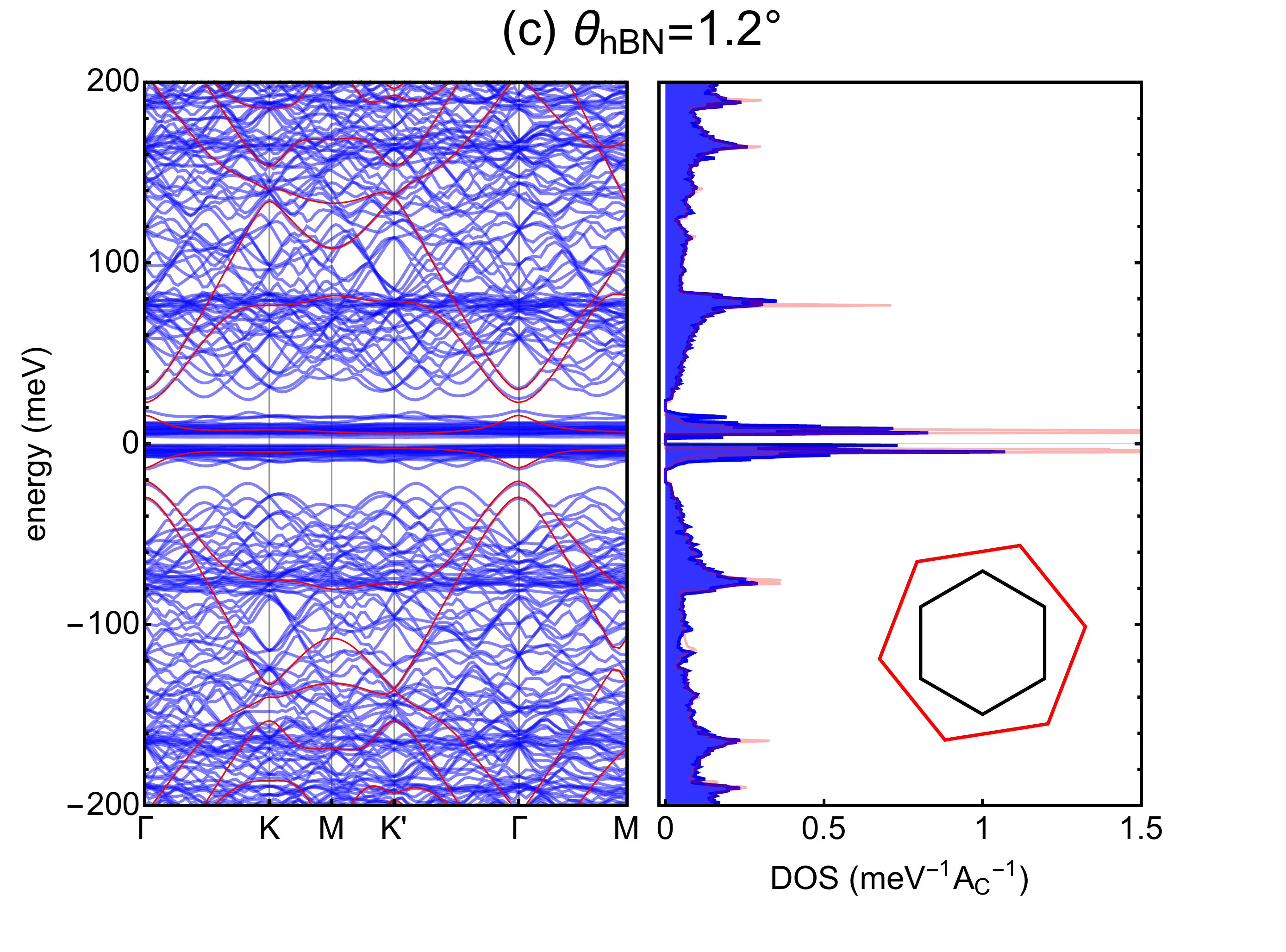}
\includegraphics[scale=0.3]{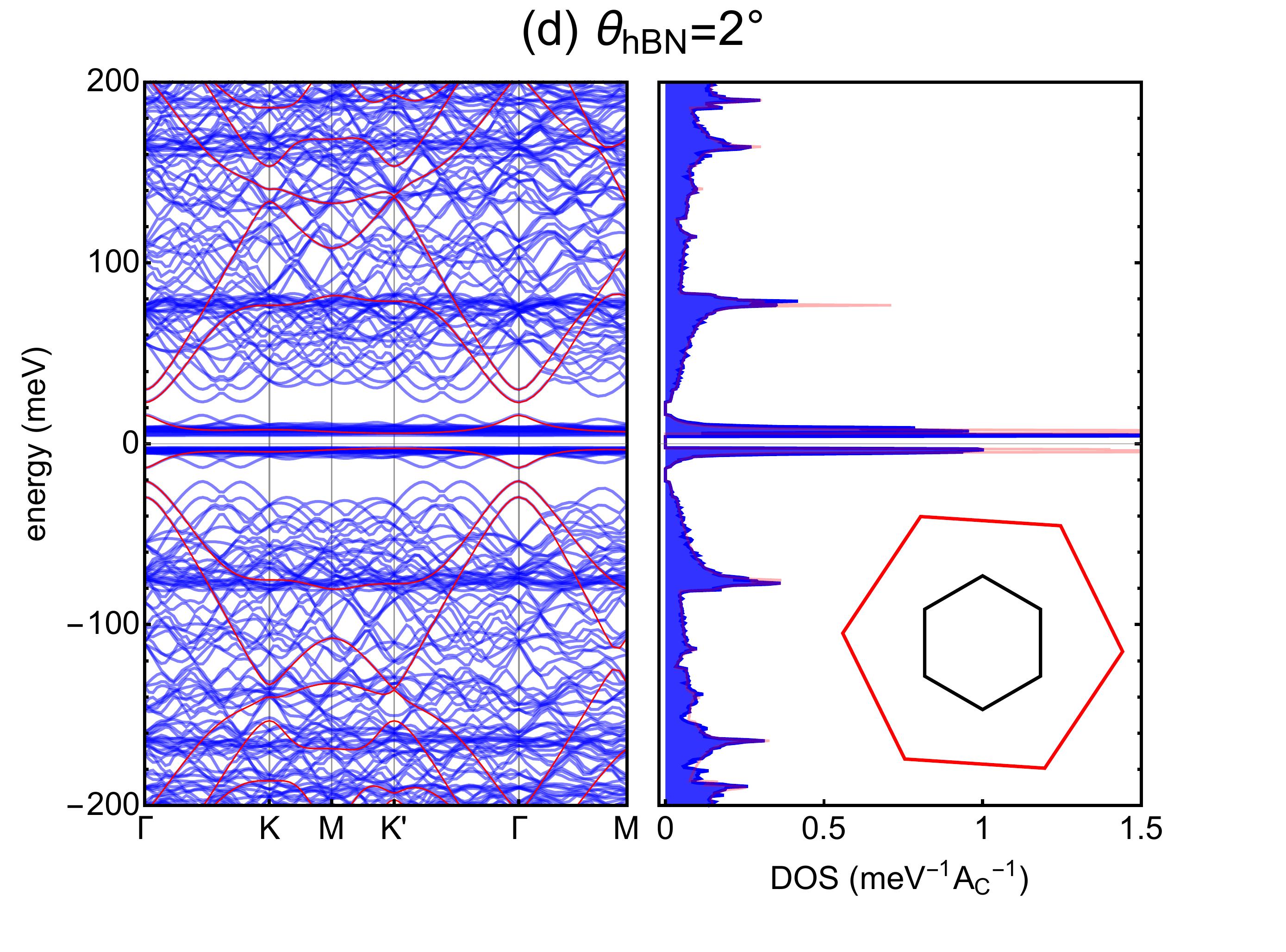}
\caption{Quasi-band structure and DOS of the non-commensurate hBN/TBG, obtained for $\Delta=40$meV.
The black and red hexagons show the two different BZs of the TBG and of the hBN/G, respectively.
The red lines refer to the band structure and the DOS of the unperturbed TBG.}
\label{non_commensurate_figs_Delta=40meV}
\end{figure*}

 The parametrization of the hBN induced potential given by the Ref.\cite{Jung2017}, Eq. \pref{vHBN_par}, accounts for a small value of the staggered potential: $\Delta=3.62$meV.
 This is the reason why the unperturbed band structure, obtained by neglecting $\delta V_{\text{SL}}(\vec{r})$ and shown by the red lines in the Fig. 5 of the main text, is almost gapless at CN and strongly resembles that of the freely standing TBG.
 To check the robustness of our findings against larger values of $\Delta$, we computed the spectrum of the non-commensurate hBN/TBG for $\Delta=40$meV.
 The results are shown in the Fig. \ref{non_commensurate_figs_Delta=40meV}, for the same values of $\theta_{hBN}$ considered in the main text.
 The red lines refer to the unperturbed spectrum, which indeed displays a sizeable gap at CN.
As is evident, for small angles the perturbed spectrum remains gapless, in very good agreement with the Figs. 5(a)-(b) of the main text.
 This means that the finite wavelengths term, $\delta V_{\text{SL}}(\vec{r})$, gives the leading contribution, rather than $\Delta$.
 This contribution becomes however negligible upon increasing $\theta_{hBN}$,
 and we recover the spectral gap already for $\theta_{hBN}\gtrsim 1^\circ$.
 
\end{document}